\documentclass[a4paper, 11pt]{article} \pdfoutput=1

\usepackage{amsmath,amssymb,amsthm,bbm,bm, amscd, mathtools, ascmac}
\usepackage{longtable, geometry}
\usepackage{enumitem}
\usepackage{latexsym}
\usepackage{braket}
\usepackage{mathrsfs}
\usepackage{authblk}
\usepackage{graphicx}
\usepackage{physics}
\usepackage{mathtools}
\usepackage{empheq}
\usepackage{stmaryrd}
\usepackage[utf8]{inputenc}
\usepackage[english]{babel}
\usepackage{esint}

\usepackage[subrefformat=parens]{subcaption}
\usepackage[svgnames,dvipsnames]{xcolor}
\usepackage[colorlinks,citecolor=DarkGreen,linkcolor=FireBrick,linktocpage,unicode]{hyperref}
\usepackage{xparse}
\usepackage{tikz}
\usetikzlibrary{arrows.meta,positioning,calc}

\usepackage{tikz-cd}
\usetikzlibrary{intersections,calc,arrows}
\usetikzlibrary{decorations.pathreplacing}
\usetikzlibrary{positioning}
\usepackage{url}
\hypersetup{urlcolor=blue}
\usepackage{comment}

\newtheorem{thm}{Theorem}[section]
\newtheorem{prop}[thm]{Proposition}
\newtheorem{lem}[thm]{Lemma}
\newtheorem{cor}[thm]{Corollary}
\theoremstyle{definition}

\newtheorem{defn}[thm]{Definition}
\newtheorem{rem}[thm]{Remark}

\newcommand{\be}{\begin{equation}}
\newcommand{\ee}{\end{equation}}

\newcommand{\im}{\mathrm{i}\,}

\numberwithin{equation}{section}

\newcommand{\bs}{\boldsymbol}

\title{
\textbf{
Stability of Kollar--Strack--Vollhardt Ferromagnetism
under Holstein and Quantized Radiation Couplings
}
}

\author[1,2,3]{Hosho Katsura}
\author[4]{Tadahiro Miyao}

\affil[1]{Department of Physics, Graduate School of Science, The University of Tokyo, Tokyo~113-0033, Japan}
\affil[2]{Institute for Physics of Intelligence, The University of Tokyo, Tokyo 113-0033, Japan}
\affil[3]{Trans-Scale Quantum Science Institute, The University of Tokyo, Tokyo 113-0033, Japan}
\affil[4]{Department of Mathematics,  Hokkaido University}

\begin{document}
\maketitle

\begin{abstract}
We study the stability of saturated ferromagnetism in the one-hole
model of Kollar, Strack, and Vollhardt (KSV) under coupling to local
Holstein phonons and, separately, to a quantized radiation field.
For Holstein phonons, with electron--phonon coupling strength $g$
and phonon frequency $\omega$, we consider the anti-adiabatic scaling
$g=g_\omega=\sqrt{E_{\rm p}\omega}$ with fixed $E_{\rm p}\ge0$.
Although the coupling may grow with $\omega$, the induced reduction
of the on-site electron--electron repulsion $U$ to $U-2E_{\rm p}$
remains finite.

Using the positive KSV decomposition and a Lang--Firsov
transformation, we obtain a lower bound for the ground-state
energy of the KSV--Holstein Hamiltonian in terms of the
ground-state energy of a scalar one-hole polaron.
This bound is attained in the fully polarized sector.
Under explicit conditions on the effective
on-site repulsion and nearest-neighbour interactions, we obtain an
exact formula for the ground-state energy and prove that every ground
state has maximum electronic spin for sufficiently large $\omega$,
with a frequency threshold independent of the system size.
We also establish stability under the Peierls coupling to a
quantized transverse radiation field, including volume-independent
conditions and, in a suitable parameter region, arbitrarily strong
electron--photon coupling.  For nonzero hopping the ground-state
space is a single maximum-spin multiplet; zero hopping permits
additional degeneracy associated with the hole position.  The other KSV ferromagnetic
branches are included, with a configuration-space connectivity
assumption in the Nagaoka-type branch. 
Both results follow from comparing the ground-state energy
of the coupled KSV system with that of a scalar one-hole model,
without assuming a volume-uniform electronic spectral gap.
\end{abstract}

%============================================================
\section{Introduction and main results}
\label{sec:introduction-main-results}
%============================================================

%------------------------------------------------------------
\subsection{Background and motivation}
\label{subsec:background-motivation}
%------------------------------------------------------------
Ferromagnetism in itinerant electron systems is one of the basic
problems in the mathematical theory of quantum many-body systems.
In localized spin models, the magnetic degrees of freedom are described
in terms of pre-existing local moments.  In itinerant-electron systems,
by contrast, ferromagnetism must emerge from the motion and interaction
of the electrons themselves, through the interplay between hopping, the
Pauli principle, and electron--electron interactions \cite{fazekas1999lecture,tasaki2020physics}.  Rigorous
examples are therefore rather delicate, especially when one asks for
stability under additional interactions.

A fundamental starting point in rigorous studies of itinerant-electron
ferromagnetism is Nagaoka's theorem, which provides a mechanism for
saturated ferromagnetism in the Hubbard model with a single hole and
infinite on-site repulsion~\cite{Nagaoka1966}.  This setting has been
studied and generalized in several directions; see, for example,
\cite{Tasaki1989,AizenmanLieb1990,Tasaki1998,KatsuraTanaka2013NagaokaSUN,Bobrow2018}.  Kollar, Strack,
and Vollhardt (KSV) introduced an extended Hubbard model with
nearest-neighbour interactions for which one-hole ferromagnetism holds
in a finite parameter region~\cite{KollarStrackVollhardt1996}.
Their argument uses a sum-of-squares decomposition and
controls charge and spin defects without requiring infinite on-site
repulsion.

The purpose of this paper is to study the stability of the KSV
ferromagnetic ground states under a local Holstein
interaction~\cite{Holstein1959}. Related stability problems have
been studied in Hubbard--Holstein models, including Nagaoka
ferromagnetism and other magnetic ground-state
structures~\cite{Miyao2017NagaokaHH,Miyao2018GroundStateHH,Miyao2019StabilityFerromagnetism}.
Here we consider the KSV extension, retaining the nearest-neighbour
interactions parameterized by $V,X,F$, and $F'$.

The local Holstein interaction is given by
$g\sum_{x\in\Lambda}n_x(b_x^*+b_x)$, where $g$ is the
electron--phonon coupling constant and $\omega$ is the phonon
frequency. We study the anti-adiabatic regime by allowing $g$
to depend on $\omega$ according to
\[
 g=g_\omega:=\sqrt{E_{\rm p}\omega},
 \qquad E_{\rm p}\ge0,
\]
with $E_{\rm p}$ fixed. Our aim is to establish the stability of one-hole ferromagnetism for
all sufficiently large $\omega$. Although $g_\omega$ grows with $\omega$ when $E_{\rm p}>0$,
the energy scale $g_\omega^2/\omega=E_{\rm p}$ associated
with the local phonon displacement remains fixed.
Consequently, the Lang--Firsov transformation
\cite{LangFirsov1963,Mahan2000} produces the nonvanishing
renormalization $U\mapsto U-2E_{\rm p}$,
where $U$ is the on-site Coulomb repulsion strength.

The main idea of our proof is to relate ferromagnetism to the
motion of a single hole coupled to local lattice vibrations.
When the electron spins are fully aligned, the motion of the
hole and its interaction with phonons are described exactly by
a scalar one-hole polaron. 
Using the Lang--Firsov transformation and the positive
decomposition of the KSV Hamiltonian, \
we show that the
ground-state energy of the KSV--Holstein Hamiltonian is
bounded below by the ground-state energy of the scalar
one-hole polaron.

This lower bound is attained by a fully polarized state
constructed from a ground state of the scalar one-hole polaron.
Moreover, under the conditions of our theorem,
the comparison shows that every ground state has no double
occupancy or local spin defects, and hence has maximum spin.
Thus, the ground-state energy is determined exactly by a
one-hole polaron problem, without requiring a spectral gap
uniform in the system size.

We also prove stability of KSV ferromagnetism when  electrons are coupled to a
quantized transverse radiation field through spin-independent
Peierls phases.
The proof uses the same positive KSV decomposition, but the 
comparison now involves a hole interacting with photons.  In
particular, the electron-photon coupling is not assumed to be small in a
suitable subregion of the electronic parameters.  The two forms
of electron-boson coupling can also be combined.

\subsection{The Kollar--Strack--Vollhardt electronic model}
\label{subsec:background-KSV-one-hole-ferromagnetism}
%------------------------------------------------------------

We begin by recalling the electronic model underlying the
KSV theorem.  Let \(\Lambda\) be a finite set of
sites, and let
\[
  \mathscr B_\Lambda
  \subset
  \bigl\{\{x,y\}:x,y\in\Lambda,\ x\neq y\bigr\}
\]
be a set of unordered bonds.  We call elements of
\(\mathscr B_\Lambda\) nearest-neighbour bonds.

For spinful fermions, let \(c_{x\sigma}\) and \(c_{x\sigma}^{*}\) be the
annihilation and creation operators at site \(x\in\Lambda\) with spin
\(\sigma\in\{\uparrow,\downarrow\}\), satisfying the canonical
anticommutation relations. Throughout, $-\sigma$ denotes the spin
opposite to $\sigma$, namely $-\uparrow=\downarrow$ and
$-\downarrow=\uparrow$. We write
\[
  n_{x\sigma}:=c_{x\sigma}^{*}c_{x\sigma},
  \qquad
  n_x:=n_{x\uparrow}+n_{x\downarrow},
\]
for the electron number operators.
The KSV nearest-neighbour Hamiltonian is
\begin{equation}\label{eq:HNN}
\begin{aligned}
  H_{\rm NN}(t,U,V,X,F,F')
  &=
  -t\sum_{\{x,y\}\in\mathscr B_\Lambda}
      \sum_{\sigma}
      \left(
        c_{x\sigma}^{*}c_{y\sigma}
        +
        c_{y\sigma}^{*}c_{x\sigma}
      \right)
  +
  U\sum_{x\in\Lambda} n_{x\uparrow}n_{x\downarrow}
  +
  V\sum_{\{x,y\}\in\mathscr B_\Lambda}
      n_x n_y
\\
  &\quad
  +
  X\sum_{\{x,y\}\in\mathscr B_\Lambda}
      \sum_{\sigma}
      \left(
        c_{x\sigma}^{*}c_{y\sigma}
        +
        c_{y\sigma}^{*}c_{x\sigma}
      \right)
      \left(
        n_{x,-\sigma}+n_{y,-\sigma}
      \right)
\\
  &\quad
  +
  F\sum_{\{x,y\}\in\mathscr B_\Lambda}
      \sum_{\sigma,\sigma'}
      c_{x\sigma}^{*}c_{y\sigma'}^{*}
      c_{x\sigma'}c_{y\sigma}
\\
  &\quad+F'\sum_{\{x,y\}\in\mathscr B_\Lambda}
      \left(
        c_{x\uparrow}^{*}c_{x\downarrow}^{*}
        c_{y\downarrow}c_{y\uparrow}
        +
        c_{y\uparrow}^{*}c_{y\downarrow}^{*}
        c_{x\downarrow}c_{x\uparrow}
      \right).
\end{aligned}
\end{equation}
Here, $t, U \in \mathbb{R}$ denote the hopping amplitude and the on-site Coulomb repulsion, respectively, while $V, X, F, F' \in \mathbb{R}$ are the nearest-neighbour coupling constants: $V$ governs the density--density interaction, $X$ the bond-charge interaction giving rise to correlated hopping, $F$ the direct exchange interaction (which is ferromagnetic when $F>0$), and $F'$ the hopping of double occupancies. 
%We note that the direct exchange interaction is ferromagnetic when $F>0$.
These extended interaction terms naturally arise when taking into account the Coulomb matrix elements involving the Wannier orbitals at nearest-neighbor sites, as originally derived by Hubbard \cite{Hubbard1963}. As emphasized by Amadon and Hirsch \cite{Amadon1996}, these extended terms are non-negligible in real materials. While the on-site repulsion $U$ is typically the dominant energy scale, the other couplings $V, X, F$, and $F'$ may have an appreciable effect on the stability of ferromagnetism.

Let
\[
  \mathcal F_\Lambda
  :=
  \bigoplus_{n=0}^{2|\Lambda|}
  \bigwedge^n
  \left(
    \ell^2(\Lambda)\otimes\mathbb C^2
  \right)
\]
be the fermionic Fock space over the spinful one-particle space $\ell^2(\Lambda)\otimes \mathbb C^2$.  The
total  electron number operator is
\[
  N_{\rm e}:=\sum_{x\in\Lambda}n_x.
\]
We work in the one-hole sector
\[
  \mathcal H_{\Lambda}^{\rm 1h}
  :=
  \operatorname{ran}
  \mathbbm 1_{\{|\Lambda|-1\}}(N_{\rm e})
  \subset\mathcal F_\Lambda,
\]
that is,
$
  N_{\rm e}=|\Lambda|-1
$
on \(\mathcal H_{\Lambda}^{\rm 1h}\).  The Hamiltonian
\(H_{\rm NN}(t,U,V,X,F,F')\) 
leaves this sector invariant, and below it is
always understood as an operator on
\(\mathcal H_{\Lambda}^{\rm 1h}\) unless otherwise stated.

We also recall the notation for the total spin. 
Let
\[
  S_x^{(j)}
  :=
  \frac12
  \sum_{\sigma,\sigma' \in\{\uparrow,\downarrow\}}
  c_{x\sigma}^{*}
  (\sigma^{(j)})_{\sigma\sigma'}
  c_{x\sigma'},
  \qquad
  j=1,2,3,
\]
where \(\sigma^{(j)}\) are the Pauli matrices.  We set
\[
  {\bm S}_x:=
  \bigl(S_x^{(1)},S_x^{(2)},S_x^{(3)}\bigr),
  \qquad
  {\bm S}_{\rm tot}:=\sum_{x\in\Lambda} {\bm S}_x.
\]
In the one-hole sector, the maximal possible total spin is
$
  S_{\max}:=(|\Lambda|-1)/2.
$
A state is said to have maximum total spin if it belongs to the
eigenspace of \({\bm S}_{\rm tot}^2\) with eigenvalue
$
  S_{\max}(S_{\max}+1).
$

%------------------------------------------------------------

%------------------------------------------------------------
\subsection{Graph assumptions and the KSV input}
\label{subsec:graph-assumptions-KSV-input}
%------------------------------------------------------------
We first formulate the graph assumptions used in the
KSV theorem~\cite{KollarStrackVollhardt1996}.
For a subset \(A\subset\Lambda\), define its edge boundary by
\[
  \partial_{\mathscr B}A
  :=
  \left\{
    \{x,y\}\in\mathscr B_\Lambda:
    x\in A,\ y\in\Lambda\setminus A
  \right\}.
\]

\begin{defn}[KSV graph condition]
\label{def:KSV-graph-condition}
We say that \((\Lambda,\mathscr B_\Lambda)\) satisfies the KSV graph
condition if the following two conditions hold.

\begin{enumerate}[label=\textup{(\roman*)}]
\item The graph $(\Lambda,\mathscr B_\Lambda)$ is finite,
connected, and $Z$-regular for an integer $Z\ge1$, meaning that
every vertex has exactly $Z$ neighbours:
\[
  \#\{y\in\Lambda:\{x,y\}\in\mathscr B_\Lambda\}=Z
  \qquad(x\in\Lambda).
\]

\item For every non-empty proper subset \(A\subsetneq\Lambda\),
\begin{equation}\label{eq:cut}
  |\partial_{\mathscr B}A|\ge Z.
\end{equation}
\end{enumerate}
\end{defn}

\begin{defn}[KSV branch]
\label{def:KSV-branch-used}
We consider the KSV branch
\[
  F>0,\qquad X\neq t.
\]
For non-bipartite graphs, we assume $t\le0$.
\end{defn}

We retain the KSV positivity function and its notation from the
electronic problem:
\[
\begin{aligned}
  \Delta_{\rm KSV}^{(1)}(t,U,V,X,F,F')
  :=U-Z\left(
    2|t|+|V-F-2|t||
    +\frac{(X-t)^2}{F}
    +\left|F'-\frac{(X-t)^2}{F}\right|
  \right).
\end{aligned}
\]

\begin{rem}[Hopping-sign convention]
\label{rem:hopping-sign-convention}
On a bipartite graph, fix a bipartition $\Lambda=A\sqcup B$ and set
\begin{equation}\label{eq:sublattice-gauge-unitary}
 G:=\exp\left\{\mathrm i\pi\sum_{x\in A}n_x\right\}.
\end{equation}
Then $G c_{x\sigma}G^*=-c_{x\sigma}$ for $x\in A$, while
$G c_{x\sigma}G^*=c_{x\sigma}$ for $x\in B$. Consequently,
\[
 G H_{\rm NN}(t,U,V,X,F,F')G^*
 =H_{\rm NN}(-t,U,V,-X,F,F').
\]
The transformation fixes every $n_x$ and acts trivially on the
phonon operators. Thus the same parameter change implements a unitary
equivalence for the KSV--Holstein Hamiltonian defined below.
Moreover, $G$ commutes with the Lang--Firsov unitary
$\mathcal U_{\rm LF}$, the total-spin operators, and the one-hole
projection. Since $|t|$ and $|X-t|$ are
invariant under $(t,X)\mapsto(-t,-X)$, so is
$\Delta_{\rm KSV}^{(1)}$.
Throughout the positive-decomposition and polaron-comparison
proofs, we use the gauge $t\le0$: for original $t>0$ on a
bipartite graph, we first replace $(t,X)$ by $(-t,-X)$ using $G$.
All displayed operator identities in these proofs, in particular
\eqref{eq:KSVexact} and \eqref{eq:intertwiner}, are understood
in this fixed gauge; the corresponding statements for the original
Hamiltonian follow by conjugation with $G^*$.
\end{rem}

\begin{thm}[Electronic KSV ground-state structure]
\label{thm:KSV-one-hole-ferromagnetism}
Assume Definitions~\ref{def:KSV-graph-condition} and
\ref{def:KSV-branch-used}.  If
$\Delta_{\rm KSV}^{(1)}(t,U,V,X,F,F')>0$, then the ground states of
$H_{\rm NN}(t,U,V,X,F,F')$ in $\mathcal H_{\Lambda}^{\rm 1h}$
have maximum total spin $S_{\max}$, and
\[
  E_{\rm GS}=-Z|t|+(V-F)(|\mathscr B_\Lambda|-Z).
\]
If $t\neq0$, the ground-state space is a single irreducible spin-$S_{\max}$
multiplet of dimension $2S_{\max}+1=|\Lambda|$; if $t=0$, it has dimension
$|\Lambda|^2$ and consists of $|\Lambda|$ such multiplets.
\end{thm}

The maximum-spin statement above is the electronic KSV result
\cite{KollarStrackVollhardt1996,KollarStrackVollhardt1997Erratum}.

\subsection{Adding a local Holstein coupling}
\label{subsec:adding-local-Holstein-coupling}
%------------------------------------------------------------
We now couple the  KSV electronic Hamiltonian to
local Holstein phonons.  We shall call the resulting model the
KSV--Holstein model.

Let
\[
  \mathcal F_{{\rm ph},\Lambda}
  :=
  \bigoplus_{n=0}^{\infty}
  \bigotimes_{\rm sym}^{n}\ell^2(\Lambda)
\]
be the bosonic Fock space over \(\ell^2(\Lambda)\).  We denote by
\(b_x^*\) and \(b_x\) the phonon creation and annihilation operators at
\(x\in\Lambda\), satisfying
\[
  [b_x,b_y^*]=\delta_{xy}\mathbbm 1,
  \qquad
  [b_x,b_y]=[b_x^*,b_y^*]=0.
\]
The phonon number operator is
\[
  N_{\rm ph}:=\sum_{x\in\Lambda}b_x^*b_x.
\]

The one-hole electron--phonon Hilbert space is
\[
  \mathcal K_{\Lambda}^{\rm 1h}
  :=
  \mathcal H_{\Lambda}^{\rm 1h}
  \otimes
  \mathcal F_{{\rm ph},\Lambda}.
\]
Throughout this paper, identity operators on tensor factors
are suppressed whenever no confusion arises. For example,
$H_{\rm NN}$ and $b_x$ denote
$H_{\rm NN}\otimes\mathbbm1$ and $\mathbbm1\otimes b_x$,
respectively.

For \(\omega>0\) and coupling \(g_\omega\), the KSV--Holstein
Hamiltonian is
\[
  H_{\rm KSVH}(\omega)
  =
  H_{\rm NN}(t,U,V,X,F,F')
  +
  \omega N_{\rm ph}
  +
  g_\omega
  \sum_{x\in\Lambda}n_x(b_x^*+b_x)
\]
on \(\mathcal K_{\Lambda}^{\rm 1h}\).  Since the electronic Hilbert
space is finite dimensional and
$
  \sum_{x\in\Lambda}n_x(b_x^*+b_x)
$
is infinitesimally \(N_{\rm ph}\)-bounded, the standard Kato--Rellich theorem~\cite[Theorem V.4.3]{Kato1995} implies that \(H_{\rm KSVH}(\omega)\) is
self-adjoint on
$
  {\rm dom}(N_{\rm ph})
$
and bounded from below.  The Holstein coupling is local and
spin-independent.  Hence \(H_{\rm KSVH}(\omega)\) commutes with the
total spin operator ${\bm S}_\mathrm{tot}$, where the spin action is understood to act trivially
on the phonon Fock factor.  Thus the notion of maximum total spin is
the same as in the electronic one-hole sector.

We work in the anti-adiabatic scaling
\[
  g_\omega=\sqrt{E_{\rm p}\omega},
  \qquad
  E_{\rm p}\ge0
\]
fixed, so that
$
  \frac{g_\omega^2}{\omega}=E_{\rm p}.
$
Thus \(E_{\rm p}\) is the polaron energy scale kept fixed as
\(\omega\to\infty\).

%------------------------------------------------------------
\subsection{Main results: direct comparison and anti-adiabatic stability}
\label{subsec:main-result-anti-adiabatic-stability}
%------------------------------------------------------------
For fixed $E_{\rm p}\ge0$, define
\[
  \Delta_{\rm KSV,\infty}^{(1)}(E_{\rm p})
  :=\Delta_{\rm KSV}^{(1)}
    (t,U-2E_{\rm p},V,X,F,F').
\]
The subscript $\infty$ refers to the anti-adiabatic limit
$\omega\to\infty$, in which the electronic Hamiltonian
$H_{\rm NN}(t,U-2E_{\rm p},V,X,F,F')$
arises naturally from the Lang--Firsov transformation.
The precise relation is given by the exact identity
\eqref{eq:exact-LF} in Section~\ref{sec:LF}.

\begin{thm}[Large-frequency KSV--Holstein stability]
\label{thm:large-omega-Kollar-Holstein-stability}
Assume the KSV graph condition and the branch and hopping-sign
convention in Definitions~\ref{def:KSV-graph-condition} and
\ref{def:KSV-branch-used}.  Fix $E_{\rm p}\ge0$ and use
$g_\omega=\sqrt{E_{\rm p}\omega}$.  If
\[
  \Delta_{\rm KSV,\infty}^{(1)}(E_{\rm p})>0,
\]
then for every sufficiently large $\omega>0$, every ground state of
$H_{\rm KSVH}(\omega)$ in $\mathcal K_{\Lambda}^{\rm 1h}$ has
maximum electronic spin $S_{\max}=(|\Lambda|-1)/2$.
More precisely, if $|t|E_{\rm p}>0$, it suffices to assume
\[
  \omega\ge\omega_0
  :=\frac{4Z|t|E_{\rm p}}
  {3\Delta_{\rm KSV,\infty}^{(1)}(E_{\rm p})}.
\]
If $|t|E_{\rm p}=0$, the same conclusion holds for every $\omega>0$.
In either case, if $t\neq0$ the ground-state space is a single
irreducible spin-$S_{\max}$ multiplet and has dimension $2S_{\max}+1=|\Lambda|$.
If $t=0$, all ground states still have maximum spin, but the ground-state
space has dimension $|\Lambda|^2$.  The frequency condition is
independent of $|\Lambda|$.
\end{thm}

\begin{rem}[Stability of all KSV ferromagnetic branches]
The large-frequency stability conclusion extends to the two remaining
KSV ferromagnetic branches under their respective strict positivity
conditions after replacing $U$ by $U-2E_{\rm p}$. The Nagaoka-type
branch additionally requires connectivity of the hole--spin
configuration graph, as specified in Appendix~\ref{app:other-KSV-branches}.

For $F>0$, $X\neq t$, this is precisely the condition
$\Delta_{\rm KSV,\infty}^{(1)}(E_{\rm p})>0$ in
Theorem~\ref{thm:large-omega-Kollar-Holstein-stability}.
For $X=t$, the corresponding condition is
\[
  U-2E_{\rm p}>
  Z\left(2|t|+\left|V-\frac F2-2|t|\right|+|F'|\right).
\]
As established in Corollary~\ref{cor:other-KSV-large-frequency}
of Appendix~\ref{app:other-KSV-branches}, stability also holds
for $F>0$, $X=t$, and for the Nagaoka-type branch
$F=0$, $X=t\neq0$, with the additional Nagaoka connectivity
condition in the latter case.

In all three branches, the stated stability holds for sufficiently
large phonon frequencies with a threshold independent of $|\Lambda|$
for fixed $Z$ and model parameters. For $t\neq0$, the ground-state
space is a single maximum-spin multiplet.
\end{rem}

\begin{rem}[Fixed electron--phonon coupling]
The stability conclusions also apply when the electron--phonon
coupling $g_\omega=g$ is kept fixed as $\omega\to\infty$.
In this case, set $E_{\rm p}(\omega)=g^2/\omega$.
The shifted KSV positivity function converges to its value at
$E_{\rm p}=0$, while Lemma~\ref{lem:lambda} gives the uniform bound
\[
 0\le\lambda_{\Lambda,\omega}
 \le Z|t|\left(1-e^{-g^2/\omega^2}\right)
 \le\frac{Z|t|g^2}{\omega^2}.
\]
Thus, if the KSV positivity condition for the corresponding
branch holds with $E_{\rm p}=0$, the stability conclusions
remain valid for all sufficiently large $\omega$, with a
volume-independent threshold under the stated graph assumptions.

The scaling $g_\omega=\sqrt{E_{\rm p}\omega}$ in the main theorem
is more demanding: it permits a nonvanishing polaron shift
$E_{\rm p}$ and hence a finite renormalization
$U\mapsto U-2E_{\rm p}$ in the large-frequency limit.
\end{rem}

\subsection{The one-hole polaron approach}
\label{subsec:direct-comparison-strategy}

In this subsection, we explain the one-hole polaron picture
underlying the proof of
Theorem~\ref{thm:large-omega-Kollar-Holstein-stability}
and formulate a finite-frequency comparison theorem
that provides the key step in the proof.

For the electronic parameters $(t,U-2E_{\rm p},V,X,F,F')$, set
\begin{equation}\label{eq:KSV-parameter-d}
  \mathfrak d:=|X-t|>0.
\end{equation}
For $a>0$, define
\begin{equation}\label{eq:KSV-tilde-coefficients}
\begin{aligned}
  \widetilde F(a)&:=F-a^2\mathfrak d,
  &\widetilde F'(a)&:=F'-\frac{\mathfrak d}{a^2},\\
  \widetilde V(a)&:=V-\frac{F+a^2\mathfrak d}{2}-2|t|,
  &\widetilde U(a)&:=U-2E_{\rm p}
  -Z\left(2|t|+|\widetilde V(a)|
    +\frac{\mathfrak d}{a^2}+|\widetilde F'(a)|\right).
\end{aligned}
\end{equation}
These are the coefficients appearing in the KSV positive decomposition
for the shifted electronic Hamiltonian.

Let $\overrightarrow{\mathscr B}_\Lambda$ contain both orientations
of each bond in $\mathscr B_\Lambda$. Set
\[
  \alpha_\omega:=\frac{g_\omega}{\omega},\qquad
  \pi_x:=\im(b_x^*-b_x),\qquad
  \mathfrak D_{xy}:=\exp\{-\im\alpha_\omega(\pi_x-\pi_y)\}.
\]
Strictly speaking, $\pi_x$ is initially defined on the
finite-phonon subspace, where it is essentially self-adjoint.
Throughout this paper, the same symbol denotes its
self-adjoint closure.

We introduce the scalar one-hole polaron Hamiltonian on
$\ell^2(\Lambda)\otimes\mathcal F_{{\rm ph},\Lambda}$ by
\[
  L_{{\rm pol},\omega}
  :=\omega N_{\rm ph}+Z|t|\mathbbm1
  -|t|\sum_{(x,y)\in\overrightarrow{\mathscr B}_\Lambda}
      |y\rangle\langle x|\otimes\mathfrak D_{xy}.
\]
The one-hole polaron energy and the reference energy are
defined, respectively, by
\begin{equation}\label{eq:polaron-reference-energies}
  \lambda_{\Lambda,\omega}:=\min\sigma(L_{{\rm pol},\omega}),
  \qquad
  E_{\rm ref}:=-Z|t|+(V-F)(|\mathscr B_\Lambda|-Z).
\end{equation}

\begin{thm}[Direct comparison with a one-hole polaron]
\label{thm:direct-KSV-Holstein-comparison}
Assume Definitions~\ref{def:KSV-graph-condition} and
\ref{def:KSV-branch-used}, fix $E_{\rm p}\ge0$ and $\omega>0$,
and use $g_\omega=\sqrt{E_{\rm p}\omega}$.
Suppose there exists $a>0$ such that
\[
  \widetilde F(a)>0,\qquad
  \lambda_{\Lambda,\omega}<\widetilde U(a).
\]
Then the ground-state energy of the KSV--Holstein Hamiltonian
is given exactly by
\begin{equation}\label{eq:energy-exact}
  \min\sigma(H_{\rm KSVH}(\omega))
  =E_{\rm ref}+\lambda_{\Lambda,\omega}
     -E_{\rm p}(|\Lambda|-1).
\end{equation}
Moreover, all ground states have maximum electronic spin $S_{\max}$.
For $t\neq0$, the ground-state space is a single spin-$S_{\max}$
multiplet; for $t=0$, its dimension is $|\Lambda|^2$.

The one-hole polaron energy satisfies
\[
  0\le\lambda_{\Lambda,\omega}
  \le Z|t|(1-e^{-E_{\rm p}/\omega}).
\]
Consequently,
\[
  Z|t|(1-e^{-E_{\rm p}/\omega})<\widetilde U(a)
\]
is a sufficient condition for the above conclusions.
\end{thm}

The large-frequency stability theorem (Theorem~\ref{thm:large-omega-Kollar-Holstein-stability}) follows from this
finite-frequency comparison. Indeed, the condition
$\Delta_{\rm KSV,\infty}^{(1)}(E_{\rm p})>0$ allows us to choose
$a>0$, independently of $\Lambda$, such that
$\widetilde F(a)>0$ and
$\widetilde U(a)\ge\tfrac34\Delta_{\rm KSV,\infty}^{(1)}(E_{\rm p})$.
Together with the upper bound on $\lambda_{\Lambda,\omega}$,
this yields the volume-independent frequency threshold in
Theorem~\ref{thm:large-omega-Kollar-Holstein-stability}.

%------------------------------------------------------------
\subsection{Stability under a quantized radiation field}
\label{subsec:main-result-radiation}
%------------------------------------------------------------
We next consider the stability of KSV ferromagnetism under
coupling to a quantized transverse radiation field through
spin-independent Peierls phases. Related lattice models
with quantized electromagnetic fields have been studied in
\cite{GiulianiMastropietroPorta2012}.
Holstein phonons are not included in this model, but
simultaneous coupling to both fields can also be treated,
as explained in
Remark~\ref{rem:simultaneous-boson-couplings}.

Let the electronic graph  $(\Lambda,\mathscr B_\Lambda)$ be embedded in a three-dimensional cubic
torus of side $L\ge L_0>0$, and let $\mathcal R_L$ be the bosonic
Fock space over
$\mathfrak h_L=\ell^2((2\pi/L)\mathbb Z^3\times\{1,2\})$.
The free radiation Hamiltonian is
\[
 H_{\rm f}=d\Gamma(\nu_L)
 =\sum_{k\in(2\pi/L)\mathbb Z^3}\sum_{\ell=1}^2
 \nu_L(k)a^*(k,\ell)a(k,\ell),
\]
where 
$\ell=1,2$ labels the two transverse photon
polarizations, $a^*(k,\ell)$ and $a(k,\ell)$ are the
corresponding creation and annihilation operators,
and
\[
 \nu_L(k)=
 \begin{cases}
 |k|,&k\neq0,\\
 m_0,&k=0,
 \end{cases}
 \qquad m_0>0.
\]
The choice $m_0>0$ is a technical convention
ensuring that the free radiation Hamiltonian has compact
resolvent at each fixed $L$.

The quantized transverse vector potential is defined by
\begin{equation}\label{eq:radiation-field}
 A_L(r)=\frac1{L^{3/2}}\sum_{k\in(2\pi/L)\mathbb Z^3}
 \sum_{\ell=1}^2\frac{\chi_\kappa(k)\varepsilon(k,\ell)}
 {\sqrt{2\nu_L(k)}}
 \bigl(e^{\im k\cdot r}a(k,\ell)
 +e^{-\im k\cdot r}a^*(k,\ell)\bigr).
\end{equation}
Here $\chi_\kappa$ is a real, even, bounded ultraviolet
cutoff function supported in $\{|k|\le\kappa\}$, where
$\kappa>0$ is fixed. For $k\neq0$, the vectors
$\varepsilon(k,1)$ and $\varepsilon(k,2)$ form a real
orthonormal basis of the plane perpendicular to $k$,
chosen so that
$\varepsilon(-k,\ell)=\varepsilon(k,\ell)$.
We set $\varepsilon(0,\ell)=0$, so that the zero-momentum
modes do not couple to the electrons.

For each oriented bond $(x,y)$, choose a piecewise smooth curve
$C_{xy}$ in the torus, with $C_{yx}$ its reversal, and set
\[
 \mathcal A_{xy}:=\int_{C_{xy}}A_L(r)\cdot dr,\qquad
 \Phi_{xy}^{\gamma}:=\exp(\im\gamma\mathcal A_{xy}),\qquad
 W_{xy}^{\gamma}:=\exp(-\im\gamma n_y\mathcal A_{xy}),
\]
where $\gamma\in\mathbb R$ is the electron--photon coupling strength.
The bond fields $\mathcal A_{xy}$ are essentially self-adjoint
on the finite-particle core, and their self-adjoint closures
are understood in the definitions of the unitaries above.

The unitary $W_{xy}^{\gamma}$ acts on the electron operators as
\[
 \begin{aligned}
 W_{xy}^{\gamma}c_{x\sigma}(W_{xy}^{\gamma})^*
 =c_{x\sigma},\qquad
 W_{xy}^{\gamma}c_{y\sigma}(W_{xy}^{\gamma})^*
 =\Phi_{xy}^{\gamma}c_{y\sigma},
 \end{aligned}
\]
while leaving all local number operators invariant. Consequently,
\[
 W_{xy}^{\gamma}c_{x\sigma}^*c_{y\sigma}
 (W_{xy}^{\gamma})^*
 =\Phi_{xy}^{\gamma}c_{x\sigma}^*c_{y\sigma}.
\]
Thus the conjugation introduces the usual Peierls phase into
electron hopping. The same phase appears in correlated hopping,
while pair hopping carries $(\Phi_{xy}^{\gamma})^2$.
The density and exchange interactions remain unchanged.

Writing
\[
 H_{\rm NN}=U\sum_xn_{x\uparrow}n_{x\downarrow}
 +\sum_{\{x,y\}\in\mathscr B_\Lambda}h_{xy},
\]
where $h_{xy}$ consists of the bond terms of \eqref{eq:HNN},
we define
\begin{equation}\label{eq:photon-Hamiltonian}
 H_{\rm KSV,\gamma}:=H_{\rm f}
 +U\sum_xn_{x\uparrow}n_{x\downarrow}
 +\sum_{\{x,y\}\in\mathscr B_\Lambda}
 W_{xy}^{\gamma}h_{xy}(W_{xy}^{\gamma})^*
\end{equation}
on $\mathcal H_\Lambda^{\rm 1h}\otimes\mathcal R_L$.
The coupling to the radiation field is spin-independent;
in particular, no direct Zeeman coupling between the electron
spins and the quantized magnetic field is included.
Consequently, $H_{\rm KSV,\gamma}$ retains its global
$\mathrm{SU}(2)$ spin symmetry.

Let $\ell^2(\Lambda)\otimes\mathcal R_L$ be the scalar one-hole
space, and define
\begin{equation}\label{eq:photon-scalar}
 \begin{split}
 L_{\Lambda,\gamma}&:=H_{\rm f}+Z|t|\mathbbm1
 -|t|\sum_{(x,y)\in\overrightarrow{\mathscr B}_\Lambda}
 |y\rangle\langle x|\otimes\Phi_{xy}^{\gamma},\\
 \lambda_{\Lambda,\gamma}&:=\min\sigma(L_{\Lambda,\gamma}).
 \end{split}
\end{equation}
The minimum exists for each finite photon box.  Write
\[
 \widetilde U_{\rm rad}(a):=U-Z\left(2|t|+|\widetilde V(a)|
 +\frac{\mathfrak d}{a^2}+|\widetilde F'(a)|\right),
 \qquad \Delta_{\rm rad}^{(1)}
 :=\Delta_{\rm KSV}^{(1)}(t,U,V,X,F,F').
\]
Here $\widetilde F(a)$, $\widetilde F'(a)$ and
$\widetilde V(a)$ are defined in
\eqref{eq:KSV-tilde-coefficients}, and
$\mathfrak d$ in \eqref{eq:KSV-parameter-d};
only the
$-2E_{\rm p}$ shift is absent from $\widetilde U_{\rm rad}(a)$.
To obtain a volume-uniform statement, suppose that the lengths
of the bond curves $C_{xy}$ are at most $\ell_*$, uniformly in the
electronic and photon boxes, and keep $\kappa,m_0,L_0$ fixed.
Define
\begin{equation}\label{eq:photon-variance}
 v_*:=\sup_{L,\Lambda}\sup_{\{x,y\}\in\mathscr B_\Lambda}
 \langle\Omega_{\rm f},\mathcal A_{xy}^2\Omega_{\rm f}\rangle,
 \qquad
 \eta_\gamma:=Z|t|(1-e^{-\gamma^2v_*/2}).
\end{equation}
The estimate $v_*<\infty$ under the above assumptions is proved
in Subsection~\ref{subsec:photon-uniform-bound}.

\begin{thm}[KSV ferromagnetism coupled to quantized radiation]
\label{thm:KSV-radiation}
Assume Definitions~\ref{def:KSV-graph-condition} and
\ref{def:KSV-branch-used}, and consider the radiation-field
Hamiltonian $H_{\rm KSV,\gamma}$ defined above.
Suppose that for some $a>0$,
\[
 \widetilde F(a)>0,\qquad
 \lambda_{\Lambda,\gamma}<\widetilde U_{\rm rad}(a).
\]
Then
\begin{equation}\label{eq:photon-energy-exact}
 \min\sigma(H_{\rm KSV,\gamma})
 =E_{\rm ref}+\lambda_{\Lambda,\gamma},
\end{equation}
and every ground state has maximum electronic spin $S_{\max}$.
For $t\ne0$ the ground-state space is a single irreducible
spin-$S_{\max}$ multiplet of dimension $2S_{\max}+1=|\Lambda|$;
for $t=0$ it has dimension $|\Lambda|^2$.

Moreover,
\begin{equation}\label{eq:photon-lambda-bounds}
 0\le\lambda_{\Lambda,\gamma}\le\eta_\gamma\le Z|t|.
\end{equation}
To relate this bound to the preceding conditions, observe that
as $a^2\uparrow F/\mathfrak d$, one has
\[
 \widetilde F(a)\downarrow0,\qquad
 \widetilde U_{\rm rad}(a)\longrightarrow
 \Delta_{\rm rad}^{(1)}.
\]
Thus, if
\begin{equation}\label{eq:photon-sufficient}
 \eta_\gamma<\Delta_{\rm rad}^{(1)},
\end{equation}
we can choose $a>0$ such that
$\widetilde F(a)>0$ and
$\lambda_{\Lambda,\gamma}<\widetilde U_{\rm rad}(a)$.
Consequently, all the above conclusions hold under
\eqref{eq:photon-sufficient}, which is independent of the
electronic and photon box sizes for fixed $Z$ and the
stated field geometry. In particular, if
$\Delta_{\rm rad}^{(1)}>Z|t|$, they hold for every
$\gamma\in\mathbb R$.
\end{thm}
\begin{cor}[Radiation stability in the $X=t$ branches]
\label{cor:KSV-radiation-other}
Assume the same graph and hopping-sign conditions, and consider
the radiation-field Hamiltonian $H_{\rm KSV,\gamma}$ defined above.
Let $X=t$ and $F\ge0$.  Put
\[
 \Delta_{\rm rad}^{(2)}:=U-Z\left(
 2|t|+\left|V-\frac F2-2|t|\right|+|F'|\right).
\]
If $\lambda_{\Lambda,\gamma}<\Delta_{\rm rad}^{(2)}$,
then \eqref{eq:photon-energy-exact} holds.  For $F>0$, every
ground state has maximum spin $S_{\rm max}$, with the multiplicities in
Theorem~\ref{thm:KSV-radiation}.  For $F=0$, the same
single-multiplet conclusion holds provided $t=X\ne0$ and
the Nagaoka connectivity condition is satisfied.
In both cases $\eta_\gamma<\Delta_{\rm rad}^{(2)}$ is a
volume-independent sufficient condition.  If
$\Delta_{\rm rad}^{(2)}>Z|t|$, no restriction on $\gamma$
is needed.
\end{cor}

The proofs of Theorem~\ref{thm:KSV-radiation} and
Corollary~\ref{cor:KSV-radiation-other} follow essentially the
same strategy as in the Holstein case.
In Appendix~\ref{app:radiation}, we outline the arguments,
focusing on the modifications required to accommodate the
quantized radiation field.

\begin{rem}[Simultaneous coupling to phonons and radiation]
\label{rem:simultaneous-boson-couplings}
The same comparison also applies when local Holstein phonons
and the quantized radiation field are present simultaneously.
After the Lang--Firsov transformation, the scalar one-hole
Hamiltonian is
\[
 L_{\omega,\gamma}=\omega N_{\rm ph}+H_{\rm f}+Z|t|\mathbbm1
 -|t|\sum_{(x,y)\in\overrightarrow{\mathscr B}_\Lambda}
 |y\rangle\langle x|\otimes\mathfrak D_{xy}
 \otimes\Phi_{xy}^{\gamma}.
\]
With $\lambda_{\Lambda,\omega,\gamma}
:=\min\sigma(L_{\omega,\gamma})$, the conclusions of
Theorem~\ref{thm:direct-KSV-Holstein-comparison} hold with
$\lambda_{\Lambda,\omega}$ replaced by
$\lambda_{\Lambda,\omega,\gamma}$, provided
$\widetilde F(a)>0$ and
$\lambda_{\Lambda,\omega,\gamma}<\widetilde U(a)$.
In particular, the exact ground-state energy is
$E_{\rm ref}-E_{\rm p}(|\Lambda|-1)
+\lambda_{\Lambda,\omega,\gamma}$, and
\[
 0\le\lambda_{\Lambda,\omega,\gamma}
 \le Z|t|\left(1-e^{-E_{\rm p}/\omega-\gamma^2v_*/2}\right).
\]
Analogous assertions hold in the $X=t$ branches, with the
additional Nagaoka connectivity condition when $F=0$.
The corresponding positivity argument on the coupled bosonic
modes is explained in Appendix~\ref{app:radiation}.
\end{rem}

%------------------------------------------------------------
\subsection{Organization of the paper}
\label{subsec:organization-of-paper}
%------------------------------------------------------------

The remainder of the paper is organized as follows.
Section~\ref{sec:LF} establishes the exact Lang--Firsov identity
and the KSV positive decomposition.
Section~\ref{sec:charge} proves that the retained kinetic
operator preserves the doublon number and derives an
energy lower bound for the no-doublon sector using the
ground-state energy of a scalar one-hole polaron.
Section~\ref{sec:ground} derives the ground-state energy
comparison and proves that all ground states have maximum spin.
Section~\ref{sec:uniqueness} determines the ground-state
multiplicity in Corollary~\ref{cor:degeneracy} and proves
the large-frequency stability result stated in
Theorem~\ref{thm:large-omega-Kollar-Holstein-stability}.
Appendix~\ref{app:CAR} provides the local CAR calculations
and the proof of the KSV decomposition.
Appendix~\ref{app:other-KSV-branches} treats the remaining
ferromagnetic branches of the KSV theorem, including the additional
configuration-space connectivity condition in the Nagaoka-type case.
The independent stability theorem for a quantized radiation field
is 
proved in Appendix~\ref{app:radiation}, which also explains the
simultaneous phonon--photon coupling.

\section{%Exact 
Lang--Firsov transformation and KSV positive decomposition}
\label{sec:LF}

For the proofs, set $N:=|\Lambda|$ and
\[
  d_x:=n_{x\uparrow}n_{x\downarrow},\qquad
  \mathscr D:=\sum_{x\in\Lambda}d_x.
\]
Here $d_x$ is the projection onto states in which site $x$
is doubly occupied by electrons of opposite spins.
Such a doubly occupied site is called a doublon, and
$\mathscr D$ counts the total number of doublons.

\subsection{%Exact 
Lang--Firsov transformation}
\label{subsec:exact-LF}

Recall $\alpha_\omega=g_\omega/\omega=\sqrt{E_{\rm p}/\omega}$ and define
\begin{equation}\label{eq:LF-unitary}
 \mathcal U_{\rm LF}=\exp\left(-\im\alpha_\omega\sum_xn_x\pi_x\right).
\end{equation}
Strictly speaking, $\sum_xn_x\pi_x$ is essentially self-adjoint
on the electronic Hilbert space tensored with the finite-phonon
subspace. Throughout this paper, unless confusion may arise, we identify
essentially self-adjoint operators with their self-adjoint closures.
By the canonical commutation relations, we have
\[
 \mathcal U_{\rm LF}b_x\mathcal U_{\rm LF}^*
 =b_x-\alpha_\omega n_x,\qquad
 \mathcal U_{\rm LF}c_{x\sigma}^*\mathcal U_{\rm LF}^*
 =c_{x\sigma}^*e^{-\im\alpha_\omega\pi_x}.
\]
The first identity shows that the transformation shifts the phonon
operators and eliminates the linear electron--phonon coupling.
Indeed, using $\omega\alpha_\omega=g_\omega$ and
$\sum_xn_x^2=N-1+2\mathscr D$ in the one-hole sector, we obtain
\begin{align}\label{eq:LF-oscillator}
 &\mathcal U_{\rm LF}\left(\omega N_{\rm ph}
 +g_\omega\sum_xn_x(b_x^*+b_x)\right)\mathcal U_{\rm LF}^*\notag\\
 &\hspace{15mm}=\omega N_{\rm ph}-E_{\rm p}\sum_xn_x^2
 =\omega N_{\rm ph}-E_{\rm p}(N-1)-2E_{\rm p}\mathscr D.
\end{align}
The second identity shows that each electron hopping term acquires
a unitary Weyl factor:
\begin{equation}\label{eq:dressed-hop}
 \mathcal U_{\rm LF}c_{x\sigma}^*c_{y\sigma}\mathcal U_{\rm LF}^*
 =c_{x\sigma}^*c_{y\sigma}\mathfrak D_{xy},
 \qquad
 \mathfrak D_{xy}=e^{-\im\alpha_\omega(\pi_x-\pi_y)},
 \quad \mathfrak D_{yx}=\mathfrak D_{xy}^*.
\end{equation}
The correlated hopping terms acquire the same Weyl factor, while
a pair hop acquires $\mathfrak D_{xy}^2$. The density and spin-exchange
terms %are 
remain unchanged.

Consequently, setting
\begin{equation}\label{eq:exact-LF}
\begin{aligned}
 H_{\rm ref}&=H_{\rm NN}(t,U-2E_{\rm p},V,X,F,F'),\\
 K_\omega&=\mathcal U_{\rm LF}H_{\rm KSVH}(\omega)
 \mathcal U_{\rm LF}^*+E_{\rm p}(N-1)\\
 &=\omega N_{\rm ph}+\mathcal U_{\rm LF}H_{\rm ref}\mathcal U_{\rm LF}^*,
\end{aligned}
\end{equation}
we obtain an exact representation of the transformed Hamiltonian.

Fix an electronic occupation configuration
$\boldsymbol n=(n_x)_x$.
On the electronic subspace with these occupation numbers,
$\mathcal U_{\rm LF}$ acts as $\mathbbm1\otimes W_{\boldsymbol n}$,
where
$
 W_{\boldsymbol n}
 :=\exp\left(-\im\alpha_\omega\sum_xn_x\pi_x\right)
$
is a Weyl displacement on $\mathcal F_{{\rm ph},\Lambda}$.
The CCR give
\[
 W_{\boldsymbol n}N_{\rm ph}W_{\boldsymbol n}^*
 =N_{\rm ph}-\alpha_\omega\sum_x n_x(b_x^*+b_x)
 +\alpha_\omega^2\sum_x n_x^2.
\]
The linear field term is infinitesimally $N_{\rm ph}$-bounded.
Thus each $W_{\boldsymbol n}$ preserves $\mathrm{dom}(N_{\rm ph})$;
since there are finitely many electronic occupation configurations,
so does $\mathcal U_{\rm LF}$. Consequently, \eqref{eq:exact-LF}
holds as an identity of self-adjoint operators.

\subsection{The retained KSV positive decomposition}\label{sec:KSV}
Put
\[
  e_x:=(1-n_{x\uparrow})(1-n_{x\downarrow}),\qquad
  p_x:=(n_{x\uparrow}-n_{x\downarrow})^2=n_x-2d_x.
\]
For later use, set
\begin{equation}\label{eq:defects}
  \mathscr S:=\sum_{\{x,y\}\in\mathscr B_\Lambda}
  \mathfrak D_{xy}^{\rm spin},\qquad
  \mathfrak D_{xy}^{\rm spin}
  :=\frac12p_xp_y-2\bs S_x\cdot\bs S_y.
\end{equation}
Both $\mathscr D$ and $\mathscr S$ are nonnegative. Indeed,
\[
  \mathfrak D_{xy}^{\rm spin}
  =\frac12p_xp_y-2\bs S_x\cdot\bs S_y
  =2P_{xy}^{\rm singlet},
\]
where $P_{xy}^{\rm singlet}$ denotes the orthogonal projection
onto the spin-singlet subspace in which both $x$ and $y$
are singly occupied ($p_x=p_y=1$), extended by zero
to all other local charge sectors.
Let
\begin{equation}\label{eq:Pis}
 \Pi_0=\mathbbm1_{\{0\}}(\mathscr D),\qquad \Pi_+=\mathbbm1-\Pi_0.
\end{equation}
Here $\Pi_0$ is the orthogonal projection onto the subspace
with no doublons, while $\Pi_+$ projects onto its orthogonal
complement. Since the electron number is $|\Lambda|-1$,
every no-doublon configuration has exactly one empty site
(a hole), with all other sites singly occupied.
These projections act on electrons and are tensored with the phonon
identity whenever needed.

Define the electronic KSV kinetic factors
\begin{equation}\label{eq:PQ}
 \begin{split}
 P_{xy\sigma}&=(1-n_{x,-\sigma})(c_{x\sigma}-c_{y\sigma})
 (1-n_{y,-\sigma}),\\
 Q_{xy\sigma}&=n_{x,-\sigma}(c_{x\sigma}-c_{y\sigma})n_{y,-\sigma},
 \end{split}
\end{equation}
and the retained nonnegative kinetic operator
\begin{equation}\label{eq:Tplus}
 \mathcal T_+=|t|\sum_{\{x,y\}\in\mathscr B_\Lambda}\sum_\sigma
 \bigl(P_{xy\sigma}P_{xy\sigma}^*+Q_{xy\sigma}^*Q_{xy\sigma}\bigr)\ge0.
\end{equation}
For the remaining KSV factors put
\begin{equation}\label{eq:AB}
 \begin{split}
 A_{xy}&=a^{-1}(c_{x\downarrow}c_{x\uparrow}
 +c_{y\downarrow}c_{y\uparrow})
 +a\operatorname{sgn}(X-t)
 (c_{y\downarrow}c_{x\uparrow}+c_{x\downarrow}c_{y\uparrow}),\\
 B_{xy}&=c_{x\downarrow}c_{x\uparrow}
 +\operatorname{sgn}(\widetilde F'(a))c_{y\downarrow}c_{y\uparrow}.
 \end{split}
\end{equation}
When $\widetilde F'(a)=0$, the sign in $B_{xy}$ is arbitrary.
Define two nonnegative diagonal  operators by
\begin{equation}\label{eq:Vplusminus}
 V_{xy}^{(+)}=e_xe_y+3d_xd_y+p_xd_y+p_yd_x,
 \qquad
 V_{xy}^{(-)}=\tfrac12(p_x-p_y)^2+2(d_xe_y+d_ye_x).
\end{equation}
For each fixed $a>0$, we choose the operator $V_{xy}$ and the
constant $v_V(a)$ according to the sign of $\widetilde V(a)$
defined in \eqref{eq:KSV-tilde-coefficients}:
\[
 V_{xy}:=
 \begin{cases}
 V_{xy}^{(+)},&\widetilde V(a)\ge0,\\
 V_{xy}^{(-)},&\widetilde V(a)<0,
 \end{cases}
 \qquad
 v_V(a):=
 \begin{cases}
 0,&\widetilde V(a)\ge0,\\
 Z/2,&\widetilde V(a)<0.
 \end{cases}
\]

\begin{lem}[Exact KSV decomposition and saturation]\label{lem:KSV}
Work in the gauge $t\le0$ and let $a>0$ satisfy
$\widetilde F(a)>0$, where $\widetilde F(a)$ is defined
in \eqref{eq:KSV-tilde-coefficients}.
On the one-hole electronic Hilbert space,
\begin{equation}\label{eq:KSVexact}
 H_{\rm ref}=E_{\rm ref}\mathbbm1+\mathcal T_++\widetilde U(a)\mathscr D+\widetilde F(a)\mathscr S+\mathcal B,
\end{equation}
where
\begin{align}\label{eq:B}
 \mathcal B={}&\mathfrak d\sum_{\{x,y\}}A_{xy}^*A_{xy}
 +|\widetilde F'(a)|\sum_{\{x,y\}}B_{xy}^*B_{xy}\notag\\
 &+|\widetilde V(a)|\left(\sum_{\{x,y\}}V_{xy}-v_V(a)\mathbbm1\right)
 +\frac{\widetilde F(a)}{2}\left((|\mathscr B_\Lambda|-Z)\mathbbm1-
 \sum_{\{x,y\}}p_xp_y\right).
\end{align}
Under \eqref{eq:cut} one has $\mathcal B\ge0$.
Moreover, $\mathcal B$, $\mathscr D$, and $\mathscr S$ annihilate
every fully spin-up one-hole state, including arbitrary
superpositions of states with different hole positions.
By spin-rotation invariance, they also annihilate the entire
maximum-total-spin subspace.
\end{lem}

The proof of Lemma~\ref{lem:KSV}, including the  CAR
calculation, positivity of $\mathcal B$, and saturation on
the maximum-spin subspace, is given in Appendix~\ref{app:CAR}.

Combining the exact Lang--Firsov identity \eqref{eq:exact-LF}
with Lemma~\ref{lem:KSV}, we obtain
\begin{equation}\label{eq:dressedKSV}
 K_\omega=E_{\rm ref}\mathbbm1+\mathcal L_\omega
 +\widetilde U(a)\mathscr D+\widetilde F(a)\mathscr S+\mathcal B_\omega,
 \qquad
 \begin{cases}
 \mathcal L_\omega=\omega N_{\rm ph}+\mathcal U_{\rm LF}\mathcal T_+\mathcal U_{\rm LF}^*\ge0,\\
 \mathcal B_\omega=\mathcal U_{\rm LF}\mathcal B\mathcal U_{\rm LF}^*\ge0.
 \end{cases}
\end{equation}
The unitary $\mathcal U_{\rm LF}$ preserves the maximum-spin subspace, and
$\mathcal B_\omega$ vanishes on that whole subspace, even when its
hole and phonon degrees of freedom are entangled.  Instead of using only the nonnegativity of $\mathcal L_\omega$,
we compare its restriction to the no-doublon sector with the
scalar one-hole polaron Hamiltonian. This allows us to retain
the polaron ground-state energy in the lower bound.

\section{Doublon conservation and scalar polaron energy comparison}
\label{sec:charge}
\subsection{Doublon conservation and reduction to the no-doublon sector}

\begin{lem}\label{lem:doublon}
Let $\mathcal T_+$ be the retained electronic kinetic operator
defined in \eqref{eq:Tplus}, and let
$\mathcal L_\omega$ be its phonon-dressed counterpart
defined in \eqref{eq:dressedKSV}.
Then
\begin{equation}\label{eq:charge-commutation}
 [\mathcal T_+,\mathscr D]=0,\qquad
 [\mathcal L_\omega,\Pi_0]=0.
\end{equation}
Moreover, the restriction of $\mathcal T_+$ to the
no-doublon one-hole subspace is given by
\begin{equation}\label{eq:Tnodoublon}
 \Pi_0\mathcal T_+\Pi_0=Z|t|\Pi_0+|t|\Pi_0
 \sum_{\{x,y\},\sigma}
 (c_{x\sigma}^*c_{y\sigma}+c_{y\sigma}^*c_{x\sigma})\Pi_0.
\end{equation}
\end{lem}

\begin{proof}
By \eqref{eq:local-PQ} and \eqref{eq:local-kinetic-identity},
the only off-diagonal terms in $\mathcal T_+$ are
\[
 \sum_\sigma(c_{x\sigma}^*c_{y\sigma}+\mathrm{h.c.})
 (1-n_{x,-\sigma}-n_{y,-\sigma})
\]
on each bond $\{x,y\}$.
For a fixed spin $\sigma$, let
$h_\sigma=c_{x\sigma}^*c_{y\sigma}$,
$a_\sigma=n_{x,-\sigma}$, and $b_\sigma=n_{y,-\sigma}$.
The CAR give
\begin{equation}\label{eq:charge-CAR}
 [d_x+d_y,h_\sigma]=h_\sigma(a_\sigma-b_\sigma),
 \qquad
 (a_\sigma-b_\sigma)(1-a_\sigma-b_\sigma)=0.
\end{equation}
Thus the off-diagonal terms commute with $d_x+d_y$.
The diagonal terms in \eqref{eq:local-PQ} also commute with
$d_x+d_y$, proving $[\mathcal T_+,\mathscr D]=0$.

On the no-doublon one-hole space, the diagonal part of
\eqref{eq:local-PQ} equals $1$ when $(n_x,n_y)=(1,0)$
or $(0,1)$, and $0$ when $(n_x,n_y)=(1,1)$.
Since exactly $Z$ bonds have one endpoint at the unique
hole, the diagonal contribution is $Z|t|\Pi_0$.
The off-diagonal part reduces to projected hopping,
giving \eqref{eq:Tnodoublon}.
Since exactly $Z$
bonds touch the unique hole, the diagonal contribution
is $Z|t|\Pi_0$. The off-diagonal part reduces to projected
hopping, giving \eqref{eq:Tnodoublon}.

Finally, $[\mathcal U_{\rm LF},\mathscr D]
=[N_{\rm ph},\mathscr D]=0$, so
$[\mathcal L_\omega,\Pi_0]=0$.
\end{proof}

The retained $\mathcal L_\omega$ is therefore block diagonal with respect to 
$\Pi_0\oplus\Pi_+$, even though $\mathcal B_\omega$ may mix these
 sectors.  Only this restricted reduction is used below.

\subsection{Comparison with the scalar one-hole polaron}\label{sec:comparison}
Fix an ordering $\Lambda=\{x_1,\ldots,x_N\}$. In the no-doublon
one-hole subspace use the basis
\begin{equation}\label{eq:hole-basis}
 |x_j,\boldsymbol\sigma\rangle
 =(-1)^{j-1}\prod_{k\ne j}^{\rightarrow}
 c_{x_k,\sigma_{x_k}}^*\Omega_{\rm el},
\end{equation}
where $\Omega_{\rm el}$ denotes the electron Fock vacuum.
Here $\prod_{k\ne j}^{\rightarrow}$ denotes the ordered product in which
the factors are arranged from left to right in increasing order
of $k$, with $k=j$ omitted.
If all spins are up this agrees with
$c_{x_j\uparrow}\prod_kc_{x_k\uparrow}^*\Omega_{\rm el}$.
For each hole position $x\in\Lambda$, define the spin space
\[
 \mathcal S_x:=\bigotimes_{y\in\Lambda\setminus\{x\}}\mathbb{C}^2.
\]
For
$\boldsymbol\sigma=(\sigma_z)_{z\in\Lambda\setminus\{x\}}
\in\{\uparrow,\downarrow\}^{\Lambda\setminus\{x\}}$,
define
\[
 |\boldsymbol\sigma\rangle_x
 :=\bigotimes_{z\in\Lambda\setminus\{x\}}|\sigma_z\rangle_z,
\]
where $|\!\uparrow\rangle_z$ and $|\!\downarrow\rangle_z$
form the standard basis of the spin space at $z$,
and the tensor factors follow the fixed ordering of $\Lambda$.
This vector corresponds to $|x,\boldsymbol\sigma\rangle$
in \eqref{eq:hole-basis}.

For adjacent sites $x,y$, define the unitary
$U_{yx}:\mathcal S_x\to\mathcal S_y$ by
\[
 U_{yx}|\boldsymbol\sigma\rangle_x
 =|\boldsymbol\sigma'\rangle_y,
\]
where
$\boldsymbol\sigma'=(\sigma'_z)_{z\in\Lambda\setminus\{y\}}
\in\{\uparrow,\downarrow\}^{\Lambda\setminus\{y\}}$ is given by
\[
 \sigma'_x=\sigma_y,\qquad
 \sigma'_z=\sigma_z\quad(z\in\Lambda\setminus\{x,y\}).
\]
Thus $U_{yx}$ transfers the spin at $y$ to $x$, leaving all
other spins unchanged. This map is unitary, with
$U_{xy}=U_{yx}^*$.

\begin{lem}[One-hole polaron representation in the no-doublon sector]
\label{lem:no-doublon-polaron}
For each hole location $h$, let $\mathcal S_h$ be the tensor product
of the $N-1$ spin-$1/2$ spaces at occupied sites. Under the
identification
\begin{equation}\label{eq:no-doublon-spin-fiber}
 \ker\mathscr D\otimes\mathcal F_{{\rm ph},\Lambda}
 \simeq\bigoplus_{h\in\Lambda}
 \left(\mathcal S_h\otimes\mathcal F_{{\rm ph},\Lambda}\right).
\end{equation}
the restriction of $\mathcal L_\omega$ to the no-doublon sector is
\begin{equation}\label{eq:Lnd}
 \mathcal L_\omega^{\rm nd}
 =\omega N_{\rm ph}+Z|t|\mathbbm1
 -|t|\sum_{(x,y)\in\overrightarrow{\mathscr B}_\Lambda}
 |y\rangle\langle x|\otimes U_{yx}\otimes\mathfrak D_{xy}.
\end{equation}

The corresponding scalar one-hole polaron Hamiltonian on
$\ell^2(\Lambda)\otimes\mathcal F_{{\rm ph},\Lambda}$ is
\begin{equation}\label{eq:Lpol}
 L_{{\rm pol},\omega}=\omega N_{\rm ph}+Z|t|\mathbbm1
 -|t|\sum_{(x,y)\in\overrightarrow{\mathscr B}_\Lambda}
 |y\rangle\langle x|\otimes\mathfrak D_{xy},
 \qquad
 \lambda_{\Lambda,\omega}=\min\sigma(L_{{\rm pol},\omega}).
\end{equation}
The hopping parts of both operators have norm at most $Z|t|$.
\end{lem}
\begin{proof}
By \eqref{eq:Tnodoublon}, the no-doublon restriction of
$\mathcal T_+$ consists of the diagonal term $Z|t|\mathbbm1$
and the projected electron hopping terms.

We first determine the fermionic sign of the hopping.
Consider an electron moving from $x_j$ into a hole at $x_i$.
With the convention in \eqref{eq:hole-basis}, the CAR give the
reordering sign $(-1)^{i+j-3}$, while the initial and final
basis phases contribute $(-1)^{i-1}$ and $(-1)^{j-1}$.
The total sign is therefore
\[
 (-1)^{(i+j-3)+(i-1)-(j-1)}=-1.
\]
Thus, writing $x=x_i$ and $y=x_j$, we obtain
\[
 \sum_\sigma c_{x\sigma}^*c_{y\sigma}
 |x,\boldsymbol\sigma\rangle
 =-|y,\boldsymbol\sigma'\rangle,
\]
where $\boldsymbol\sigma'$ is obtained by transferring the spin
at $y$ to $x$, leaving all other spins unchanged.
This transfer defines the unitary $U_{yx}:\mathcal S_x\to\mathcal S_y$.
In particular, the hopping sign is independent of the spin
configuration. Combining this with the Weyl dressing in
\eqref{eq:dressed-hop} gives \eqref{eq:Lnd}.
Omitting the spin transport yields the scalar operator
\eqref{eq:Lpol}.

Finally, let $\Psi=(\Psi_x)_{x\in\Lambda}$ be a vector in
the direct sum \eqref{eq:no-doublon-spin-fiber}, where
$\Psi_x\in\mathcal S_x\otimes\mathcal F_{{\rm ph},\Lambda}$
is the component corresponding to the hole position $x$.
By the unitarity of $U_{yx}$ and $\mathfrak D_{xy}$,
together with the Cauchy--Schwarz inequality,
\[
 \begin{aligned}
 \sum_{y\in\Lambda}
 \left\|\sum_{x\sim y}U_{yx}\mathfrak D_{xy}\Psi_x\right\|^2
 \le Z\sum_{y\in\Lambda}\sum_{x\sim y}\|\Psi_x\|^2
 =Z^2\sum_{x\in\Lambda}\|\Psi_x\|^2.
 \end{aligned}
\]
Here we used the $Z$-regularity of the graph.
Thus the hopping operator in \eqref{eq:Lnd} has norm at most
$Z|t|$. The same bound holds for \eqref{eq:Lpol}.
\end{proof}

\begin{lem}[Lower bound by the scalar polaron energy]\label{lem:modulus}
For every $\omega>0$ and every
\[
 \Psi=(\Psi_x)_{x\in\Lambda}\in
 \bigoplus_{x\in\Lambda}
 \left(\mathcal S_x\otimes\mathrm{dom}(N_{\rm ph}^{1/2})\right),
\]
there exists a nonnegative vector
$\rho\in\ell^2(\Lambda)\otimes\mathrm{dom}(N_{\rm ph}^{1/2})$,
with $\|\rho\|=\|\Psi\|$, such that
\begin{equation}\label{eq:modulus}
 \mathfrak q_{\mathcal L_\omega^{\rm nd}}[\Psi]
 \ge\mathfrak q_{L_{{\rm pol},\omega}}[\rho]
 \ge\lambda_{\Lambda,\omega}\|\Psi\|^2.
\end{equation}
Here $\mathfrak q_T$ denotes the closed quadratic form of the
nonnegative self-adjoint operator $T$.
\end{lem}
\begin{proof}
Identify $\mathcal F_{{\rm ph},\Lambda}$ with $L^2(\mathbb{R}^N,dq)$, with
\begin{equation}\label{eq:osc-coordinate}
 N_{\rm ph}=\tfrac12\sum_{z\in\Lambda}
 (-\partial_{q_z}^2+q_z^2-1),\qquad
 (\mathfrak D_{xy}\phi)(q)=\phi(q+\boldsymbol{\xi}_{xy}),\quad
 \boldsymbol{\xi}_{xy}=\sqrt2\alpha_\omega(\mathbf e_y-\mathbf e_x),
\end{equation}
where $\mathbf e_z$ denotes the standard unit vector of
$\mathbb{R}^N$ corresponding to $z\in\Lambda$.
Identifying $\mathcal S_h\otimes L^2(\mathbb R^N,dq)$
with $L^2(\mathbb R^N,dq;\mathcal S_h)$,
write $\Psi=(\Psi_h)_{h\in\Lambda}$, where
$\Psi_h\in L^2(\mathbb R^N,dq;\mathcal S_h)$, and put
\begin{equation}\label{eq:rho}
 \rho_h(q)=\|\Psi_h(q)\|_{\mathcal S_h}.
\end{equation}
The oscillator form domain is
\begin{equation}\label{eq:form-domain}
 \mathcal Q=\left\{\Psi:
 \Psi_h\in H^1(\mathbb{R}^N;\mathcal S_h),\quad
 |q|\Psi_h\in L^2\text{ for every }h\right\}.
\end{equation}
Here $H^1(\mathbb{R}^N;\mathcal S_h)$ denotes the Sobolev space
of $\mathcal S_h$-valued square-integrable functions whose
first weak derivatives are also square-integrable.
The finite direct sum is implicit. The norm map on a Hilbert space
is $1$-Lipschitz, and the Sobolev chain rule yields
$\rho_h\in H^1$ and
\[
 |\partial_{q_z}\rho_h(q)|
 \le\|\partial_{q_z}\Psi_h(q)\|_{\mathcal S_h}
\]
for almost every $q\in\mathbb R^N$.
Moreover, $|q|\rho_h\in L^2$ and $\|\rho\|=\|\Psi\|$.
Thus $\rho$ belongs to the form domain of
$L_{{\rm pol},\omega}$.
 Since
$\rho_h(q)^2=\|\Psi_h(q)\|_{\mathcal S_h}^2$,
the multiplication terms $q_z^2-1$ contribute equally to the
oscillator quadratic forms for $\rho$ and $\Psi$, while the
gradient terms do not increase. Therefore,
\begin{equation}\label{eq:osc-modulus}
 \|N_{\rm ph}^{1/2}\rho\|^2
 \le\|N_{\rm ph}^{1/2}\Psi\|^2.
\end{equation}
For a directed hopping term, $U_{yx}$ is unitary and $\mathfrak
D_{xy}$ is a real translation, so pointwise Cauchy--Schwarz gives
\begin{align}\label{eq:hop-CS}
 &\operatorname{Re}\left\langle\Psi_y(q),
 U_{yx}\Psi_x(q+\boldsymbol{\xi}_{xy})\right\rangle_{\mathcal S_y}
 \le\rho_y(q)\rho_x(q+\boldsymbol{\xi}_{xy}).
\end{align}
The hopping coefficient is $-|t|\le0$; hence the corresponding
energy inequality has the required direction. Combining the
hopping bound with \eqref{eq:osc-modulus} gives
\[
 \mathfrak q_{\mathcal L_\omega^{\rm nd}}[\Psi]
 \ge\mathfrak q_{L_{{\rm pol},\omega}}[\rho].
\]
The lower bound by $\lambda_{\Lambda,\omega}\|\rho\|^2$
follows from the variational characterization of the bottom of the
spectrum.
\end{proof}

\begin{lem}[Ground-state existence and energy bounds for the scalar one-hole polaron]\label{lem:lambda}
For $\omega>0$, the operator $L_{{\rm pol},\omega}$ is self-adjoint,
has compact resolvent at fixed finite volume, and satisfies
\begin{equation}\label{eq:lambda}
 0\le\lambda_{\Lambda,\omega}
 \le Z|t|(1-e^{-E_{\rm p}/\omega})
 \le\frac{Z|t|E_{\rm p}}{\omega}.
\end{equation}
In particular, its lowest spectral point is an eigenvalue.
\end{lem}
\begin{proof}
Since $\mathfrak D_{yx}=\mathfrak D_{xy}^*$ and
$\|\mathfrak D_{xy}\|=1$, the hopping term in \eqref{eq:Lpol}
is a bounded self-adjoint operator. Thus $L_{{\rm pol},\omega}$
is self-adjoint on $\mathrm{dom}(N_{\rm ph})$.
Since $N_{\rm ph}$ has compact resolvent for fixed finite
$\Lambda$, so does $L_{{\rm pol},\omega}$.

Using the identification
$
 \ell^2(\Lambda)\otimes\mathcal F_{{\rm ph},\Lambda}
 \simeq\bigoplus_{h\in\Lambda}\mathcal F_{{\rm ph},\Lambda},
$
write $\phi=(\phi_h)_{h\in\Lambda}$, where
$\phi_h\in\mathcal F_{{\rm ph},\Lambda}$.
The unitarity of $\mathfrak D_{xy}$ and the
$Z$-regularity of the graph give
\begin{equation}\label{eq:polaron-squares}
 \left\langle\phi,
 \left(Z|t|\mathbbm1-|t|
 \sum_{(x,y)\in\overrightarrow{\mathscr B}_\Lambda}
 |y\rangle\langle x|\otimes\mathfrak D_{xy}\right)\phi
 \right\rangle
 =|t|\sum_{\{x,y\}\in\mathscr B_\Lambda}
 \|\phi_y-\mathfrak D_{xy}\phi_x\|^2\ge0.
\end{equation}
Here either orientation may be chosen for each bond on the
right-hand side. Consequently, $L_{{\rm pol},\omega}\ge0$.

For the normalized trial state
$
 \phi=\frac1{\sqrt N}\sum_{h\in\Lambda}
 |h\rangle\otimes\Omega_{\rm ph},
$
the Weyl vacuum expectation is
\[
 \langle\Omega_{\rm ph},
 \mathfrak D_{xy}\Omega_{\rm ph}\rangle
 =e^{-\alpha_\omega^2}=e^{-E_{\rm p}/\omega}.
\]
Since there are $NZ$ directed bonds, we obtain
\[
 \langle\phi,L_{{\rm pol},\omega}\phi\rangle
 =Z|t|(1-e^{-E_{\rm p}/\omega}).
\]
The variational principle and $1-e^{-s}\le s$ for $s\ge0$
give \eqref{eq:lambda}.
\end{proof}

\section{Ground-state comparison and spin identification}\label{sec:ground}
\begin{prop}[Global form inequality and exact energy]\label{prop:energy}
Under $\widetilde F(a)>0$, $\widetilde U(a)>0$, the quadratic-form inequality
\begin{equation}\label{eq:energy-inequality}
 K_\omega-(E_{\rm ref}+\lambda_{\Lambda,\omega})\mathbbm1
 \ge(\widetilde U(a)-\lambda_{\Lambda,\omega})\mathscr D+\widetilde F(a)\mathscr S
\end{equation}
holds for every $\omega>0$.  If $\lambda_{\Lambda,\omega}<\widetilde U(a)$,
then the exact ground-state energy formula \eqref{eq:energy-exact}
holds and every ground state lies in
$\ker\mathscr D\cap\ker\mathscr S$.
\end{prop}
\begin{proof}
By Lemma~\ref{lem:doublon}, $\mathcal L_\omega$ is reduced by the
orthogonal decomposition $\mathrm{ran}\,\Pi_0\oplus\mathrm{ran}\,\Pi_+$.  By Lemma~\ref{lem:modulus}, its
no-doublon restriction is bounded below by
$\lambda=\lambda_{\Lambda,\omega}$; its other block is nonnegative by \eqref{eq:dressedKSV}.
Thus
\begin{equation}\label{eq:L-lower}
 \mathcal L_\omega-\lambda\Pi_0\ge0.
\end{equation}
By \eqref{eq:dressedKSV}, the \emph{exact} difference between the
two sides of \eqref{eq:energy-inequality} is
\begin{align}\label{eq:exact-positive-difference}
 &K_\omega-(E_{\rm ref}+\lambda)\mathbbm1
 -(\widetilde U(a)-\lambda)\mathscr D-\widetilde F(a)\mathscr S\notag\\
 &\qquad=(\mathcal L_\omega-\lambda\Pi_0)
 +\mathcal B_\omega+\lambda(\mathscr D-\Pi_+)\ge0.
\end{align}
The last term is nonnegative. Indeed, $\mathscr D$ counts the
number of doublons and therefore has nonnegative integer
eigenvalues. It vanishes on $\operatorname{ran}\Pi_0$ and is
at least one on $\operatorname{ran}\Pi_+$. Hence
$\mathscr D\ge\Pi_+$.

Choose a ground state $\Phi$ of $L_{{\rm pol},\omega}$, and
let $\Psi_\uparrow$ denote its embedding into the fully
spin-up one-hole subspace.
Since $\mathcal U_{\rm LF}$ preserves this subspace,
Lemma~\ref{lem:KSV} gives
$
 \mathcal B_\omega\Psi_\uparrow
 =\mathscr D\Psi_\uparrow
 =\mathscr S\Psi_\uparrow=0.
$
Moreover, under the same embedding,
$\mathcal L_\omega\Psi_\uparrow
=\lambda_{\Lambda,\omega}\Psi_\uparrow$.
Thus, by \eqref{eq:dressedKSV},
\[
 K_\omega\Phi=(E_{\rm ref}+\lambda)\Phi.
\]
If $\widetilde U(a)>\lambda$, the lower bound
\eqref{eq:energy-inequality} and $\widetilde F(a)>0$ show that
$E_{\rm ref}+\lambda$ is the ground-state energy of $K_\omega$.
For any ground vector $\Psi$, taking the expectation of
\eqref{eq:energy-inequality} gives
\[
 0\ge(\widetilde U(a)-\lambda)
 \langle\Psi,\mathscr D\Psi\rangle
 +\widetilde F(a)\langle\Psi,\mathscr S\Psi\rangle\ge0.
\]
Since both coefficients are strictly positive and
$\mathscr D,\mathscr S\ge0$, it follows that
$\mathscr D\Psi=\mathscr S\Psi=0$.
Finally, the scalar shift in \eqref{eq:exact-LF} gives
\eqref{eq:energy-exact}.
\end{proof}

\begin{lem}[Connectivity after removing the hole]\label{lem:hole-graph}
Assume that $(\Lambda,\mathscr B_\Lambda)$ is a finite,
connected, $Z$-regular graph with $Z\ge1$ satisfying
\eqref{eq:cut}. Then the induced graph on
$\Lambda\setminus\{h\}$ is connected for every $h\in\Lambda$.
\end{lem}
\begin{proof}
If the induced graph on $\Lambda\setminus\{h\}$ had two or more
nonempty connected components, then by \eqref{eq:cut} at least
$Z$ edges would leave each component. Since no edges join
distinct components, all such edges must be incident to $h$.
This would give at least $2Z$ distinct edges incident to $h$,
contradicting $\deg(h)=Z$.
\end{proof}

\begin{prop}[All ground states have maximum spin]\label{prop:spin}
If $\lambda_{\Lambda,\omega}<\widetilde U(a)$ and $\widetilde F(a)>0$, every ground state of
$H_{\rm KSVH}(\omega)$ has total electronic spin $S=S_{\max}=(N-1)/2$.
\end{prop}

\begin{proof}
By Proposition~\ref{prop:energy}, every ground vector $\Psi$
satisfies
$
 \mathscr D\Psi=0
 $
 and 
 $
  \mathscr S\Psi=0.
$
Since each $\mathfrak D_{xy}^{\rm spin}$ is nonnegative and
$\mathscr S=\sum_{\{x,y\}}\mathfrak D_{xy}^{\rm spin}$, we have
\[
 \mathfrak D_{xy}^{\rm spin}\Psi=0
 \qquad(\{x,y\}\in\mathscr B_\Lambda).
\]

The absence of doublons implies that each electronic configuration
has exactly one hole. We may therefore decompose $\Psi$ according
to the hole position $h\in\Lambda$. For each $h$, the corresponding
spin-phonon component $\Psi_h$ belongs to
$
\mathcal S_h
 \otimes\mathcal F_{{\rm ph},\Lambda}.
$
On this subspace, for any bond $\{x,y\}$ with $x,y\neq h$,
\[
 \mathfrak D_{xy}^{\rm spin}
 =\frac12-2\bs S_x\cdot\bs S_y
 =1-P_{xy}^{\rm spin},
\]
where $P_{xy}^{\rm spin}$ is the unitary operator that
exchanges the spin states at $x$ and $y$, leaving all
other spins unchanged. Consequently,
\[
 (P_{xy}^{\rm spin}\otimes\mathbbm1)\Psi_h=\Psi_h
 \qquad(x,y\neq h,\ \{x,y\}\in\mathscr B_\Lambda).
\]
Thus $\Psi_h$ is invariant under the interchange of spins at
any two adjacent occupied sites.

By Lemma~\ref{lem:hole-graph}, the graph obtained by removing
$h$ is connected. The transpositions associated with the edges
of a connected graph generate the full symmetric group on its
vertices. Hence $\Psi_h$ is invariant under every permutation
of the $N-1$ spin factors. It follows that
\[
 \Psi_h\in
 \left(\otimes_{\rm sym}^{N-1}\mathbb{C}^2\right)
 \otimes\mathcal F_{{\rm ph},\Lambda}.
\]
The completely symmetric tensor space
$\otimes_{\rm sym}^{N-1}\mathbb{C}^2$ carries the irreducible
spin-$S_{\max}$ representation, where $S_{\max}=(N-1)/2$.
Since this holds for every hole position $h$, the ground vector
$\Psi$ belongs to the maximum-total-spin eigenspace.

Finally, the Lang--Firsov unitary commutes with the total-spin
operators. Therefore the same conclusion holds for every ground
state of the original KSV--Holstein Hamiltonian.
\end{proof}

\section{Ground-state multiplicity and the stability theorem}\label{sec:uniqueness}
\subsection{Uniqueness of the scalar-polaron ground state}
The preceding comparison shows that all ground states have maximum
spin but does not by itself exclude multiple copies of the
spin-$S_{\max}$ representation. We now prove scalar-polaron
ground-state uniqueness without assuming connectivity of the full
hole--spin configuration graph.

\begin{prop}[Positivity improvement]\label{prop:PF}
If $t\neq0$ and $(\Lambda,\mathscr B_\Lambda)$ is connected, then for every $\beta>0$ the
scalar-polaron semigroup $e^{-\beta L_{{\rm pol},\omega}}$ is positivity
improving on the cone
\begin{equation}\label{eq:cone}
 \mathscr P=\{(\phi_h)_h:\phi_h(q)\ge0
 \text{ for almost every }q\text{ and every }h\}
 \subset\ell^2(\Lambda)\otimes L^2(\mathbb R^N).
\end{equation}
Consequently
\begin{equation}\label{eq:polaron-simple}
 \dim\ker(L_{{\rm pol},\omega}-\lambda_{\Lambda,\omega})=1.
\end{equation}
\end{prop}
\begin{proof}
Write $L_{{\rm pol},\omega}=A+Z|t|\mathbbm1-J$, with
\begin{equation}\label{eq:PF-AJ}
 A=\omega N_{\rm ph},\qquad
 J=|t|\sum_{(x,y)\in\overrightarrow{\mathscr B}_\Lambda}
 |y\rangle\langle x|\otimes\mathfrak D_{xy},\qquad
 \|J\|\le Z|t|.
\end{equation}
The Weyl factors are real translations in phonon coordinates,
so $J$ preserves $\mathscr P$.  For every $s>0$, $e^{-sA}$
acts separately in each hole component by the harmonic-oscillator
Mehler kernel.  This kernel is strictly positive for every pair of
phonon coordinates; hence it sends every nonzero positive function
in a given hole component to a strictly positive one.

The bounded-perturbation Dyson expansion is
\begin{align}\label{eq:Dyson}
 e^{-\beta L_{{\rm pol},\omega}}
 &=e^{-\beta Z|t|}\sum_{m=0}^\infty
 \int_{0<s_1<\cdots<s_m<\beta}
 e^{-(\beta-s_m)A}J e^{-(s_m-s_{m-1})A}
 \cdots J e^{-s_1A}\,
 \mathrm{d}s_1\cdots\mathrm{d}s_m.
\end{align}
For $m=0$ the integral means $e^{-\beta A}$.  Every term is
positivity preserving, and the series converges in operator norm
because the $m$-th term has norm at most
$\beta^m\|J\|^m/m!$.

Take nonzero $\varphi,\psi\in\mathscr P$.  Choose $x,y$ with
$\psi_x\ne0$ and $\varphi_y\ne0$.  As the graph is connected, there is a path
$x=x_0\sim x_1\sim\cdots\sim x_m=y$.
In the $m$-th Dyson term, select the product of hopping operators
corresponding to this path.  For all times in the interior of the
integration simplex, the positive Mehler kernels, followed by
coordinate translations, produce a strictly positive function in
hole component $y$ from the nonzero $\psi_x$.  Its inner product with
$\varphi_y\ge0$, $\varphi_y\ne0$, is strictly positive. Since $|t|>0$,
this path contributes positively, while every other term is
nonnegative.  Thus
\begin{equation}\label{eq:improving}
 \langle\varphi,e^{-\beta L_{{\rm pol},\omega}}\psi\rangle>0
 \quad(0\ne\varphi,\psi\in\mathscr P),
\end{equation}
which is positivity improvement. Here the cone is self-dual in
the complex Hilbert-space sense:
\[
 \mathscr P=\left\{f\in\ell^2(\Lambda)\otimes L^2(\mathbb R^N):
 \langle f,g\rangle\in[0,\infty)
 \text{ for every }g\in\mathscr P\right\}.
\]
Indeed, testing against nonnegative functions in a single hole
component forces each component of $f$ to be real and nonnegative
almost everywhere. By Lemma~\ref{lem:lambda},
$L_{{\rm pol},\omega}$ is self-adjoint, bounded below, and has compact
resolvent, so $\lambda_{\Lambda,\omega}$ is an eigenvalue.
The Perron--Frobenius--Faris simplicity criterion for a
positivity-improving self-adjoint semigroup therefore gives
\eqref{eq:polaron-simple}; see \cite{Faris1972} and the related
semigroup method in \cite{Miyao2017NagaokaHH}.
\end{proof}

\subsection{The entire ground-state multiplet}\label{sec:multiplet}

\begin{prop}[Unitary equivalence in the maximum-spin sector]
\label{prop:intertwiner}
In the gauge $t\le0$, let
$\mathcal H_{S=S_{\max}}\subset\mathcal H_{\Lambda}^{\rm 1h}$
denote the spin-$S_{\max}$ eigenspace of ${\bs S}_{\rm tot}^2$,
where $S_{\max}=(N-1)/2$. There is a unitary
\[
 W:\bigl(\ell^2(\Lambda)\otimes\mathcal F_{{\rm ph},\Lambda}\bigr)
 \otimes\mathbb{C}^{2S_{\max}+1}
 \longrightarrow\mathcal H_{S=S_{\max}}\otimes\mathcal F_{{\rm ph},\Lambda}
\]
such that
\begin{equation}\label{eq:intertwiner}
 W^*(K_\omega\!\upharpoonright_{S=S_{\max}})W
 =(E_{\rm ref}\mathbbm1+L_{{\rm pol},\omega})
 \otimes\mathbbm1_{2S_{\max}+1}.
\end{equation}
\end{prop}
\begin{proof}
Write $S_{\rm tot}^-:=\sum_xc_{x\downarrow}^*c_{x\uparrow}$.
For each hole $h$ put
$|h,S_{\max}\rangle=c_{h\uparrow}\prod_xc_{x\uparrow}^*\Omega_{\rm el}$.
Let $n=N-1=2S_{\max}$. For $k=0,\ldots,n$ define
\begin{equation}\label{eq:spin-weight}
 |h,S_{\max}-k\rangle
 =\sqrt{\frac{(n-k)!}{k!\,n!}}\,
 (S_{\rm tot}^-)^k|h,S_{\max}\rangle.
\end{equation}
For each $h$, these form the standard orthonormal basis of the
irreducible spin-$S_{\max}$ representation. Moreover, an electron
state with $r\ge1$ doublons has at most $n-2r$ singly occupied
sites and hence total spin at most $(n-2r)/2<S_{\max}$.
Thus the entire spin-$S_{\max}$ electronic subspace is the
direct sum of $N$ symmetric spin subspaces, one for each
hole position.
Define
\[
 W(|h\rangle\otimes\phi\otimes|S_{\max}-k\rangle)
 =|h,S_{\max}-k\rangle\otimes\phi.
\]
This extends to a unitary.

On the highest-weight subspace, Lemma~\ref{lem:KSV} and
\eqref{eq:dressedKSV} give exactly
$K_\omega=E_{\rm ref}+L_{{\rm pol},\omega}$.
To extend this identity to all spin weights in
\eqref{eq:spin-weight}, we establish
$[K_\omega,S_{\rm tot}^-]=0$.
We first verify the $\mathrm{SU}(2)$ invariance of $H_{\rm ref}$. In particular, the
correlated-hopping operator is a spin scalar: by
\eqref{eq:local-singlets} it equals
$C_{xy}=R_{xy}^*M_{xy}+M_{xy}^*R_{xy}$, where $R_{xy}$ and
$M_{xy}$ are singlet-pair annihilators. The ordinary 
hopping, the density and pair-hopping terms, and the exchange term
\eqref{eq:exchange-CAR} are spin invariant as well. Since
$\mathcal U_{\rm LF}$ is generated by the local charge operators
$n_x$ and phonon operators, it commutes with the total-spin
generators. Hence $[K_\omega,S_{\rm tot}^-]=0$ on
$\mathrm{dom}(N_{\rm ph})$. Since $[K_\omega,S_{\rm tot}^-]=0$ and the normalization in
\eqref{eq:spin-weight} is independent of the hole position,
the identity obtained on the highest-weight subspace extends
to every $k=0,\ldots,2S_{\max}$.
More precisely, under the identification defined by $W$,
the restriction of $K_\omega$ to the subspace with
$S_{\rm tot}^{(3)}=S_{\max}-k$ within the maximum-spin sector
is given by $E_{\rm ref}\mathbbm1+L_{{\rm pol},\omega}$.

Since $H_{\rm ref}$ is bounded on the finite-dimensional electronic
space, \eqref{eq:exact-LF} implies that $K_\omega$ is self-adjoint
on the electronic space tensored with $\mathrm{dom}(N_{\rm ph})$.
Similarly, $L_{{\rm pol},\omega}$ is self-adjoint on
$\ell^2(\Lambda)\otimes\mathrm{dom}(N_{\rm ph})$.
The unitary $W$ leaves the phonon factor unchanged, and hence
\[
 W\left(\mathrm{dom}(L_{{\rm pol},\omega})
 \otimes\mathbb C^{2S_{\max}+1}\right)
 =\mathrm{dom}(K_\omega\!\upharpoonright_{S=S_{\max}}).
\]
Thus \eqref{eq:intertwiner} is an identity of self-adjoint
operators, including their domains.
For an original $t>0$ bipartite model, the corresponding intertwiner
is $G^*W$, where $W$ is constructed above for the gauge-transformed
Hamiltonian. Here $G$ is the sublattice unitary defined in
\eqref{eq:sublattice-gauge-unitary}.
\end{proof}

\begin{cor}[Precise degeneracy]\label{cor:degeneracy}
Under the hypotheses of Proposition~\ref{prop:spin},
\begin{equation}\label{eq:ker-factorization}
 \ker(K_\omega-E_{\rm ref}-\lambda_{\Lambda,\omega})
 =W\Bigl(\ker(L_{{\rm pol},\omega}-\lambda_{\Lambda,\omega})
 \otimes\mathbb{C}^{2S_{\max}+1}\Bigr).
\end{equation}
If $|t|>0$, its dimension is $N$. If $|t|=0$, then
$L_{{\rm pol},\omega}=\omega N_{\rm ph}$ on each of the $N$ hole components,
and the dimension is $N\cdot N=N^2$.
\end{cor}
\begin{proof}
By Proposition~\ref{prop:spin}, all ground states belong to
$S=S_{\max}$. Proposition~\ref{prop:intertwiner} then gives the
factorization. For $|t|>0$ use Proposition~\ref{prop:PF}.
For $|t|=0$, the phonon vacuum is the unique ground state
in each independent hole component.
\end{proof}

\subsection{Explicit volume-independent frequency condition}\label{sec:frequency}

The following lemma explains the role of the KSV positivity condition
in comparing the ground-state energies of $H_{\rm KSVH}(\omega)$
and $L_{{\rm pol},\omega}$.
For $0<\theta<1$, define $a_\theta>0$ by
\begin{equation}\label{eq:a-theta}
 a_\theta^2=\frac{(1-\theta)F}{\mathfrak d}.
\end{equation}
Then $\widetilde F(a_\theta)=\theta F>0$, while
\[
 \lim_{\theta\downarrow0}\widetilde U(a_\theta)
 =\Delta_{\rm KSV,\infty}^{(1)}(E_{\rm p}).
\]
Thus, if the KSV positivity function is strictly positive, we can
choose $a$ so that both $\widetilde F(a)$ and $\widetilde U(a)$ are
positive. The lemma makes this choice quantitative and uniform
in the volume.

\begin{lem}[Uniform choice of the KSV parameter]\label{lem:a-choice}
If $\Delta_{\rm KSV,\infty}^{(1)}(E_{\rm p})>0$, there is an $a>0$, depending only on
$F,X,t,U,V,F',E_{\rm p},Z$, such that
\begin{equation}\label{eq:uf}
 \widetilde F(a)>0,\qquad \widetilde U(a)\ge\frac34\Delta_{\rm KSV,\infty}^{(1)}(E_{\rm p}).
\end{equation}
\end{lem}
\begin{proof}
Write $\Delta=\Delta_{\rm KSV,\infty}^{(1)}(E_{\rm p})>0$,
$c=\mathfrak d^2/F$, and $w=V-F-2|t|$.
For $0<\theta\le1/2$, take $a=a_\theta$ as defined in
\eqref{eq:a-theta}. Then
\begin{equation}\label{eq:theta-params}
\begin{aligned}
 \widetilde F(a)&=\theta F,
 &\widetilde V(a)&=w+\frac{\theta F}{2},\\
 \frac{\mathfrak d}{a^2}&=\frac{c}{1-\theta},
 &\widetilde F'(a)&=F'-\frac{c}{1-\theta}.
\end{aligned}
\end{equation}
Using $||r|-|s||\le|r-s|$ in the definition of
$\widetilde U(a)$, we obtain
\begin{equation}\label{eq:U-parameter-theta}
 |\widetilde U(a)-\Delta|
 \le Z\left(\frac{\theta F}{2}
 +\frac{2c\theta}{1-\theta}\right)
 \le Z\theta\left(\frac F2+4c\right).
\end{equation}
Now choose
\begin{equation}\label{eq:theta-choice}
 \theta=\min\left\{\frac14,\,
 \frac{\Delta}{4Z(F/2+4c)}\right\}>0.
\end{equation}
Then \eqref{eq:U-parameter-theta} gives
\[
 \widetilde U(a)\ge\frac34\Delta>0,
 \qquad \widetilde F(a)=\theta F>0.
\]
This proves \eqref{eq:uf}. Since the choice of $\theta$ depends
only on the model parameters and $Z$, it is independent of
$|\Lambda|$.
\end{proof}

\subsection{Proof of Theorems \ref{thm:large-omega-Kollar-Holstein-stability} and \ref{thm:direct-KSV-Holstein-comparison} }

\begin{proof}[Proof of Theorem~\ref{thm:direct-KSV-Holstein-comparison}]
By Lemma~\ref{lem:lambda}, we have
$\lambda_{\Lambda,\omega}\ge0$, so the hypothesis
$\lambda_{\Lambda,\omega}<\widetilde U(a)$ implies
$\widetilde U(a)>0$.
The exact ground-state energy follows from
Proposition~\ref{prop:energy}.
The maximum-spin assertion follows from
Proposition~\ref{prop:spin}, and the ground-state multiplicities
from Corollary~\ref{cor:degeneracy}.
The stated bound on $\lambda_{\Lambda,\omega}$ and the sufficient
finite-frequency condition follow from Lemma~\ref{lem:lambda}.
\end{proof}

\begin{proof}[Proof of Theorem~\ref{thm:large-omega-Kollar-Holstein-stability}]
Write $\Delta:=\Delta_{\rm KSV,\infty}^{(1)}(E_{\rm p})>0$.
Choose $a$ as in Lemma~\ref{lem:a-choice}, so that
$\widetilde U(a)\ge3\Delta/4$ and $\widetilde F(a)>0$.
Suppose first that $|t|E_{\rm p}>0$ and $\omega\ge\omega_0$.
Set
\[
 r:=\frac{3\Delta}{4Z|t|}>0.
\]
By the definition of $\omega_0$, we have $E_{\rm p}/\omega\le r$.
The exponential bound in Lemma~\ref{lem:lambda} therefore gives
\begin{equation}\label{eq:frequency-check}
\begin{aligned}
 \widetilde U(a)-\lambda_{\Lambda,\omega}
 &\ge \frac{3\Delta}{4}
   -Z|t|\bigl(1-e^{-E_{\rm p}/\omega}\bigr)\\
 &\ge \frac{3\Delta}{4}
   -Z|t|\bigl(1-e^{-r}\bigr)\\
 &=Z|t|\bigl(r-1+e^{-r}\bigr)
 =:\delta_0>0.
\end{aligned}
\end{equation}
The strict positivity follows from $1-e^{-r}<r$ for $r>0$.
In particular, the estimate holds also at $\omega=\omega_0$,
and $\delta_0$ is independent of $|\Lambda|$.

If $|t|E_{\rm p}=0$, Lemma~\ref{lem:lambda} gives
$\lambda_{\Lambda,\omega}=0$ for every $\omega>0$. Hence
\[
 \widetilde U(a)-\lambda_{\Lambda,\omega}
 \ge\frac{3\Delta}{4}>0.
\]
In either case, the assertions follow from
Theorem~\ref{thm:direct-KSV-Holstein-comparison}
and Corollary~\ref{cor:degeneracy}.
\end{proof}

\appendix

\section{Proof of the KSV decomposition and saturation}
\label{app:CAR}

We give the local CAR identities used in the proof of
Lemma~\ref{lem:KSV} and establish the positivity and saturation
properties of the KSV decomposition. We also record the local
identity needed in Lemma~\ref{lem:doublon}.
Several of these identities appear in the original KSV papers
\cite{KollarStrackVollhardt1996,KollarStrackVollhardt1997Erratum};
we reproduce the relevant calculations for the reader's convenience.

\subsection{Local CAR identities}

Fix a bond $\{x,y\}$ and write
\begin{align*}
 T_{xy}&=\sum_\sigma(c_{x\sigma}^*c_{y\sigma}+\mathrm{h.c.}),\\
 C_{xy}&=\sum_\sigma(c_{x\sigma}^*c_{y\sigma}+\mathrm{h.c.})
 (n_{x,-\sigma}+n_{y,-\sigma}),\\
 J_{xy}&=c_{x\uparrow}^*c_{x\downarrow}^*c_{y\downarrow}c_{y\uparrow}
 +\mathrm{h.c.}
\end{align*}
Using the definitions of $P,Q,A,B$ in \eqref{eq:PQ} and
\eqref{eq:AB}, the CAR give
\begin{align}
 \sum_\sigma(P_{xy\sigma}P_{xy\sigma}^*
 +Q_{xy\sigma}^*Q_{xy\sigma})
 &=4\mathbbm1-3(n_x+n_y)+2(d_x+d_y)
 +2n_xn_y+T_{xy}-C_{xy},
 \label{eq:local-PQ}\\
 \mathfrak d A_{xy}^*A_{xy}
 &=\frac{\mathfrak d}{a^2}(d_x+d_y+J_{xy})
 +a^2\mathfrak d\left(\frac12n_xn_y-2\bs S_x\cdot\bs S_y\right)
 +(X-t)C_{xy},
 \label{eq:local-A}\\
 |\widetilde F'(a)|B_{xy}^*B_{xy}
 &=|\widetilde F'(a)|(d_x+d_y)+\widetilde F'(a)J_{xy}.
 \label{eq:local-B}
\end{align}

For completeness, the terms in \eqref{eq:local-A} can be
verified by introducing
\[
 R_{xy}=c_{x\downarrow}c_{x\uparrow}
 +c_{y\downarrow}c_{y\uparrow},\qquad
 M_{xy}=c_{y\downarrow}c_{x\uparrow}
 +c_{x\downarrow}c_{y\uparrow}.
\]
Direct calculation gives
\begin{equation}\label{eq:local-singlets}
\begin{aligned}
 R_{xy}^*R_{xy}&=d_x+d_y+J_{xy},\\
 M_{xy}^*M_{xy}&=\frac12n_xn_y-2\bs S_x\cdot\bs S_y,\\
 R_{xy}^*M_{xy}+M_{xy}^*R_{xy}&=C_{xy}.
\end{aligned}
\end{equation}
The exchange term in \eqref{eq:HNN} satisfies
\begin{equation}\label{eq:exchange-CAR}
\begin{aligned}
 \sum_{\sigma,\sigma'}c_{x\sigma}^*c_{y\sigma'}^*
 c_{x\sigma'}c_{y\sigma}
 &=-\sum_{\sigma,\sigma'}
 (c_{x\sigma}^*c_{x\sigma'})
 (c_{y\sigma'}^*c_{y\sigma})\\
 &=-\frac12n_xn_y-2\bs S_x\cdot\bs S_y.
\end{aligned}
\end{equation}

For the diagonal operators, the identities
$e_x=\mathbbm1-n_x+d_x$ and $p_x=n_x-2d_x$ yield
\begin{equation}\label{eq:local-V}
\begin{aligned}
 V_{xy}^{(+)}
 &=\mathbbm1-n_x-n_y+d_x+d_y+n_xn_y,\\
 V_{xy}^{(-)}
 &=\frac12(n_x+n_y)+d_x+d_y-n_xn_y.
\end{aligned}
\end{equation}
On the one-hole space,
\[
 \sum_{\{x,y\}}(n_x+n_y)=Z(N-1),\qquad
 \sum_{\{x,y\}}(d_x+d_y)=Z\mathscr D,
\]
and $2|\mathscr B_\Lambda|=ZN$. Consequently,
\[
\begin{aligned}
 \sum_{\{x,y\}}V_{xy}^{(+)}
 &=Z\mathscr D+\sum_{\{x,y\}}n_xn_y
 -(|\mathscr B_\Lambda|-Z)\mathbbm1,\\
 \sum_{\{x,y\}}V_{xy}^{(-)}-\frac Z2\mathbbm1
 &=Z\mathscr D-\sum_{\{x,y\}}n_xn_y
 +(|\mathscr B_\Lambda|-Z)\mathbbm1.
\end{aligned}
\]
Thus, for either sign of $\widetilde V(a)$,
\begin{align}\label{eq:V-summed}
 |\widetilde V(a)|
 \left(\sum_{\{x,y\}}V_{xy}-v_V(a)\mathbbm1\right)
 =Z|\widetilde V(a)|\mathscr D
 +\widetilde V(a)\sum_{\{x,y\}}n_xn_y
 -\widetilde V(a)(|\mathscr B_\Lambda|-Z)\mathbbm1.
\end{align}
Finally, the definition of $\mathscr S$ gives
\begin{align}\label{eq:spin-p-p}
 &\widetilde F(a)\mathscr S
 +\frac{\widetilde F(a)}2
 \left((|\mathscr B_\Lambda|-Z)\mathbbm1
 -\sum_{\{x,y\}}p_xp_y\right)\notag\\
 &\quad=-2\widetilde F(a)
 \sum_{\{x,y\}}\bs S_x\cdot\bs S_y
 +\frac{\widetilde F(a)}2
 (|\mathscr B_\Lambda|-Z)\mathbbm1.
\end{align}

We also record an equivalent form of the hopping contribution
in \eqref{eq:local-PQ}:
\begin{equation}\label{eq:local-kinetic-identity}
 T_{xy}-C_{xy}
 =\sum_\sigma(c_{x\sigma}^*c_{y\sigma}+\mathrm{h.c.})
 (1-n_{x,-\sigma}-n_{y,-\sigma}).
\end{equation}
This identity will be used to prove the doublon-number
conservation in Lemma~\ref{lem:doublon}.

\subsection{Proof of Lemma~\ref{lem:KSV}}

\begin{proof}[Proof of Lemma~\ref{lem:KSV}]
We first establish the exact decomposition \eqref{eq:KSVexact}.
Expanding its right-hand side by
\eqref{eq:local-PQ}--\eqref{eq:local-B},
\eqref{eq:V-summed}, and \eqref{eq:spin-p-p},
we obtain the following coefficients:
\[
\begin{aligned}
 \mathscr D:\quad&
 \widetilde U(a)+Z\left(2|t|+|\widetilde V(a)|
 +\frac{\mathfrak d}{a^2}+|\widetilde F'(a)|\right)
 =U-2E_{\rm p},\\
 T_{xy}:\quad&|t|=-t,\\
 C_{xy}:\quad&-|t|+(X-t)=X,\\
 J_{xy}:\quad&\frac{\mathfrak d}{a^2}+\widetilde F'(a)=F',\\
 \bs S_x\cdot\bs S_y:\quad&
 -2(a^2\mathfrak d+\widetilde F(a))=-2F,\\
 n_xn_y:\quad&
 2|t|+\frac12a^2\mathfrak d+\widetilde V(a)=V-\frac F2.
\end{aligned}
\]
By \eqref{eq:exchange-CAR}, the last two coefficients
together reproduce the density--density and exchange
interactions in \eqref{eq:HNN}.
The remaining scalar is
\[
\begin{aligned}
 &E_{\rm ref}+|t|[4|\mathscr B_\Lambda|-3Z(N-1)]
 +\left(\frac{\widetilde F(a)}2-\widetilde V(a)\right)
 (|\mathscr B_\Lambda|-Z)\\
 &=E_{\rm ref}+Z|t|+(F-V)(|\mathscr B_\Lambda|-Z)=0,
\end{aligned}
\]
where we used
$\widetilde F(a)/2-\widetilde V(a)=F-V+2|t|$
and the definition of $E_{\rm ref}$ in
\eqref{eq:polaron-reference-energies}.
This proves \eqref{eq:KSVexact}.

We next prove $\mathcal B\ge0$.
The first two terms in \eqref{eq:B} are nonnegative. %squares.
Since
\[
 \sum_xe_x=\sum_x(\mathbbm1-n_x+d_x)
 =\mathscr D+\mathbbm1,
\]
every charge configuration in the one-hole sector contains
at least one empty site. The $Z$ bonds incident to such a site
do not contribute to $\sum_{\{x,y\}}p_xp_y$. Hence
\begin{equation}\label{eq:Npp}
 \sum_{\{x,y\}}p_xp_y
 \le (|\mathscr B_\Lambda|-Z)\mathbbm1.
\end{equation}
Since $\widetilde F(a)>0$, the last term in \eqref{eq:B}
is nonnegative.

For $\widetilde V(a)\ge0$, each $V_{xy}^{(+)}$ is nonnegative
and $v_V(a)=0$.
Suppose $\widetilde V(a)<0$. In a fixed charge configuration,
let $E,P,D$ be the sets of empty, singly occupied, and doubly
occupied sites, respectively. From the definition of
$V_{xy}^{(-)}$,
\begin{equation}\label{eq:Vminus-cut}
 \sum_{\{x,y\}}V_{xy}^{(-)}
 =\frac12|\partial_{\mathscr B}P|+2b(D,E),
\end{equation}
where $b(D,E)$ counts the bonds connecting $D$ to $E$.
If $P\neq\varnothing$, then $P$ is a proper subset of $\Lambda$
because an empty site exists. Thus \eqref{eq:cut} gives
$|\partial_{\mathscr B}P|\ge Z$.
If $P=\varnothing$, the one-hole condition implies that
$D$ and $E$ are both nonempty, and
$b(D,E)=|\partial_{\mathscr B}D|\ge Z$.
Therefore,
\[
 \sum_{\{x,y\}}V_{xy}^{(-)}\ge\frac Z2\mathbbm1.
\]
Since $v_V(a)=Z/2$ in this case, the third term in
\eqref{eq:B} is also nonnegative. We conclude that
$\mathcal B\ge0$.

It remains to prove the saturation statement.
For a completely spin-up one-hole state with the hole at $h$,
there are no doublons or spin singlets, and hence
$\mathscr D=\mathscr S=0$ on this state.
Moreover, $A_{xy}$ and $B_{xy}$ annihilate it because
they contain down-spin annihilation operators.
For every hole position $h$, direct evaluation gives
\begin{equation}\label{eq:saturation}
 \sum_{\{x,y\}}V_{xy}^{(+)}=0,\qquad
 \sum_{\{x,y\}}V_{xy}^{(-)}=\frac Z2\mathbbm1,\qquad
 \sum_{\{x,y\}}p_xp_y
 =(|\mathscr B_\Lambda|-Z)\mathbbm1
\end{equation}
on the completely spin-up one-hole subspace.
It follows from \eqref{eq:B} that
$\mathcal B$, $\mathscr D$, and $\mathscr S$ annihilate
the entire completely spin-up one-hole subspace,
including arbitrary superpositions of states with
different hole positions.

Finally, these three operators commute with global spin
rotations. Since the maximum-spin subspace is generated
from completely spin-up one-hole states by repeated application
of $S_{\rm tot}^-$, the same conclusion holds on the entire
maximum-total-spin subspace.
\end{proof}

\section{The other ferromagnetic branches of the KSV theorem}
\label{app:other-KSV-branches}

We extend the direct comparison to the two remaining KSV branches
with $X=t$. Throughout this appendix we assume the KSV graph
condition in Definition~\ref{def:KSV-graph-condition} and use the
hopping-sign convention $t\le0$ from
Remark~\ref{rem:hopping-sign-convention}.
For an original $t>0$ bipartite model, we first apply the unitary
$G$ defined in that remark, which sends $(t,X)$ to $(-t,-X)$.
The branch condition $X\ne t$
from Definition~\ref{def:KSV-branch-used} is not imposed here.

\subsection{The KSV decomposition when $X=t$}

For $X=t$ and $F\ge0$, set
\begin{equation}\label{eq:other-KSV-coefficients}
 \widetilde V^{(2)}:=V-\frac F2-2|t|,\qquad
 \widetilde U^{(2)}:=U-2E_{\rm p}
 -Z\bigl(2|t|+|\widetilde V^{(2)}|+|F'|\bigr).
\end{equation}
These are the KSV coefficients corresponding to Cases~1(b) and~2
in the classification of Kollar, Strack, and
Vollhardt~\cite{KollarStrackVollhardt1996}, evaluated at the
shifted electronic parameters $(t,U-2E_{\rm p},V,t,F,F')$.
Unlike the branch $X\neq t$, no auxiliary parameter $a$ is needed.
The reference energy is still
\[
 E_{\rm ref}=-Z|t|+(V-F)(|\mathscr B_\Lambda|-Z).
\]
Let
\[
 B^{(2)}_{xy}:=c_{x\downarrow}c_{x\uparrow}
 +\operatorname{sgn}(F')c_{y\downarrow}c_{y\uparrow},
\]
where the sign is arbitrary when $F'=0$.  Put
\[
 (V_{xy}^{(2)},v_2):=
 \begin{cases}
 (V_{xy}^{(+)},0),&\widetilde V^{(2)}\ge0,\\
 (V_{xy}^{(-)},Z/2),&\widetilde V^{(2)}<0,
 \end{cases}
\]
with $V_{xy}^{(\pm)}$ as in \eqref{eq:Vplusminus}, and define
\begin{align}\label{eq:other-KSV-remainder}
 \mathcal B^{(2)}:={}&|F'|\sum_{\{x,y\}\in\mathscr B_\Lambda}
 (B^{(2)}_{xy})^*B^{(2)}_{xy}\notag
 +|\widetilde V^{(2)}|\left(
 \sum_{\{x,y\}\in\mathscr B_\Lambda}V_{xy}^{(2)}-v_2\mathbbm1\right)\notag\\
 &+\frac F2\left((|\mathscr B_\Lambda|-Z)\mathbbm1
 -\sum_{\{x,y\}\in\mathscr B_\Lambda}p_xp_y\right).
\end{align}

\begin{lem}[Positive decomposition for $X=t$]
\label{lem:other-KSV-decomposition}
On $\mathcal H_\Lambda^{\rm 1h}$,
\begin{equation}\label{eq:other-KSV-identity}
 H_{\rm ref}=E_{\rm ref}\mathbbm1+\mathcal T_+
 +\widetilde U^{(2)}\mathscr D+F\mathscr S+\mathcal B^{(2)},
 \qquad \mathcal B^{(2)}\ge0.
\end{equation}
Moreover, the entire no-doublon one-hole space is contained in
$\ker\mathcal B^{(2)}$; in particular,
\begin{equation}\label{eq:other-KSV-remainder-kernel}
 \mathcal B^{(2)}\Pi_0=\Pi_0\mathcal B^{(2)}=0.
\end{equation}
\end{lem}
\begin{proof}
In the local CAR identities \eqref{eq:local-PQ}--\eqref{eq:local-B}
set $\mathfrak d=|X-t|=0$.  The $A_{xy}^*A_{xy}$ term disappears,
$\widetilde F=F$, $\widetilde F'=F'$, and
$\widetilde V=\widetilde V^{(2)}$.
The coefficient calculation and scalar normalization in
Appendix~\ref{app:CAR} give \eqref{eq:other-KSV-identity}.

The square terms in \eqref{eq:other-KSV-remainder} are positive.
The one-hole inequality \eqref{eq:Npp} makes the last term
nonnegative.  The diagonal comparison for $V_{xy}^{(+)}$ and
$V_{xy}^{(-)}$, proved in Lemma~\ref{lem:KSV} using
\eqref{eq:cut}, proves $\mathcal B^{(2)}\ge0$.

On the no-doublon one-hole space, every $B^{(2)}_{xy}$ vanishes,
and the unique hole implies
\[
 \sum_{\{x,y\}}V_{xy}^{(+)}=0,\qquad
 \sum_{\{x,y\}}V_{xy}^{(-)}=\frac Z2\mathbbm1,\qquad
 \sum_{\{x,y\}}p_xp_y=(|\mathscr B_\Lambda|-Z)\mathbbm1.
\]
Thus $\mathcal B^{(2)}\Pi_0=0$.
Taking adjoints and using the self-adjointness of
$\mathcal B^{(2)}$ and $\Pi_0$, we obtain
$\Pi_0\mathcal B^{(2)}=0$, completing
\eqref{eq:other-KSV-remainder-kernel}.
\end{proof}

Write $\mathcal B^{(2)}_\omega=\mathcal U_{\rm LF}
\mathcal B^{(2)}\mathcal U_{\rm LF}^*$.  Since
$[\mathcal U_{\rm LF},\Pi_0]=0$, Lemma~\ref{lem:other-KSV-decomposition}
and the  Lang--Firsov identity yield
\begin{equation}\label{eq:other-dressed-KSV}
 K_\omega=E_{\rm ref}\mathbbm1+\mathcal L_\omega
 +\widetilde U^{(2)}\mathscr D+F\mathscr S
 +\mathcal B^{(2)}_\omega,
 \qquad \mathcal B^{(2)}_\omega\ge0,
 \quad \mathcal B^{(2)}_\omega\Pi_0=0.
\end{equation}
Here $\mathcal L_\omega$ is the same retained operator as in
\eqref{eq:dressedKSV}.  Since $\mathcal L_\omega$ in \eqref{eq:other-dressed-KSV}
is identical to that in \eqref{eq:dressedKSV},
Lemmas~\ref{lem:doublon} and \ref{lem:modulus}
apply also to the present case.
In particular, the operator inequality \eqref{eq:L-lower}
remains valid.
Consequently,
\begin{equation}\label{eq:other-global-comparison}
 K_\omega-(E_{\rm ref}+\lambda_{\Lambda,\omega})\mathbbm1
 \ge(\widetilde U^{(2)}-\lambda_{\Lambda,\omega})\mathscr D
 +F\mathscr S.
\end{equation}
Indeed, the difference of the two sides equals
\[
 (\mathcal L_\omega-\lambda_{\Lambda,\omega}\Pi_0)
 +\mathcal B^{(2)}_\omega
 +\lambda_{\Lambda,\omega}(\mathscr D-\Pi_+)\ge0.
\]
The ground state of the scalar-polaron embedded in the fully polarized
 space of one-hole reaches equality in \eqref{eq:other-global-comparison}.

\subsection{Case 1(b): $F>0$ and $X=t$}

\begin{prop}[Finite-frequency stability in Case 1(b)]
\label{prop:other-KSV-case1b}
Suppose $F>0$, $X=t$, and
\begin{equation}\label{eq:other-case1b-condition}
 \widetilde U^{(2)}>\lambda_{\Lambda,\omega}.
\end{equation}
Then
\[
 \min\sigma(H_{\rm KSVH}(\omega))
 =E_{\rm ref}+\lambda_{\Lambda,\omega}-E_{\rm p}(|\Lambda|-1),
\]
and every ground state has total spin $S_{\max}$.  If $t\ne0$,
the ground-state space is a single spin-$S_{\max}$ multiplet;
if $t=0$, its dimension is $|\Lambda|^2$.
\end{prop}
\begin{proof}
By \eqref{eq:other-global-comparison} and $F>0$, every ground state
is annihilated by $\mathscr D$ and $\mathscr S$.  The sharpness of
the lower bound gives the stated energy.  The connectivity of
$\Lambda\setminus\{h\}$ from Lemma~\ref{lem:hole-graph} shows that
$\ker\mathscr D\cap\ker\mathscr S$ belongs to the maximum-spin sector.
The scalar-polaron uniqueness argument in Proposition~\ref{prop:PF}
and the maximum-spin intertwiner in Proposition~\ref{prop:intertwiner}
apply since $\mathcal B^{(2)}_\omega$ annihilates the whole
maximum-spin sector.  They give the stated multiplicities.
\end{proof}

\subsection{Case 2: $F=0$ and $X=t\ne0$}

We formulate explicitly the additional connectivity assumption used
in the Nagaoka-type branch. Fix an eigenvalue $M$ of
$S_{\rm tot}^{(3)}$, and let $\mathscr C_M$ be the set of one-hole
spin configurations $(h,\boldsymbol\sigma)$ satisfying
\[
 S_{\rm tot}^{(3)}|h,\boldsymbol\sigma\rangle
 =M|h,\boldsymbol\sigma\rangle.
\]
Here $h$ denotes the hole position and
$\boldsymbol\sigma$ the spin configuration on
$\Lambda\setminus\{h\}$. The allowed values are
$M=-S_{\max},-S_{\max}+1,\ldots,S_{\max}$.
Two configurations are adjacent if the hole can move across one bond
while the electron at the other endpoint is transported into the hole.
We require the configuration graph to be connected for every allowed
$M$ (the \emph{Nagaoka connectivity condition}; see
\cite{Tasaki1998}). This condition is not implied by
the KSV cut condition \eqref{eq:cut}; see
\cite{KollarStrackVollhardt1996,Miyao2017NagaokaHH}.

\begin{lem}[Positivity improvement in a fixed-magnetization sector]
\label{lem:other-KSV-fixed-M-PF}
Assume $t\ne0$, $\omega>0$, and the Nagaoka connectivity condition.
For every allowed eigenvalue $M$ of $S_{\rm tot}^{(3)}$, the
restriction of $\mathcal L_\omega^{\rm nd}$ to the fixed-$M$ sector
has compact resolvent and generates a positivity-improving semigroup
with respect to the cone of nonnegative phonon-coordinate functions
in the hole--spin configuration basis. In particular, its lowest
eigenvalue is simple.
\end{lem}
\begin{proof}
Use the basis \eqref{eq:hole-basis} and the phonon-coordinate
representation \eqref{eq:osc-coordinate}. Identifying the fixed-$M$
sector with
\[
 \mathcal H_M=\bigoplus_{c\in\mathscr C_M}
 L^2(\mathbb R^N,dq),
\]
we may write its Hamiltonian as
\[
 \mathcal L_{\omega,M}^{\rm nd}
 =A+Z|t|\mathbbm1-J_M,\qquad A=\omega N_{\rm ph},
\]
where, writing $h(c)$ for the hole position of a configuration $c$,
\[
 J_M=|t|\sum_{c\to c'}|c'\rangle\langle c|
 \otimes\mathfrak D_{h(c),h(c')}.
\]
The sum is over directed edges of the configuration graph. The
fermionic sign is $-1$ for every allowed hole move, by the CAR
calculation in Lemma~\ref{lem:no-doublon-polaron}; thus the
coefficient of $J_M$ in the Hamiltonian is negative, independent
of the spin configuration. Reversing a move replaces its Weyl
operator by its adjoint. Moreover, each configuration has exactly
$Z$ possible hole moves. Consequently, $J_M=J_M^*$, and Cauchy--Schwarz gives
\[
\begin{aligned}
 \|J_M\psi\|^2
 &\le Z|t|^2\sum_{c\in\mathscr C_M}
 \sum_{c'\sim c}\|\psi_{c'}\|^2
 =Z^2|t|^2\|\psi\|^2.
\end{aligned}
\]
In particular, $\|J_M\|\le Z|t|$.
Thus $\mathcal L_{\omega,M}^{\rm nd}$ is self-adjoint on the
finite direct sum of copies of $\mathrm{dom}(N_{\rm ph})$ and has
compact resolvent, because $N_{\rm ph}$ has compact resolvent at
finite volume and $J_M$ is bounded.

Define the cone
\[
 \mathscr P_M=\left\{\psi=(\psi_c)_{c\in\mathscr C_M}:
 \psi_c(q)\ge0\ \text{for a.e. }q\in\mathbb R^N
 \text{ and every }c\right\}.
\]
It is self-dual in the standard complex Hilbert-space sense.
By \eqref{eq:osc-coordinate}, each $\mathfrak D_{xy}$ acts as a
real translation and therefore preserves nonnegativity. Hence
$J_M\mathscr P_M\subseteq\mathscr P_M$.
For $s>0$, the semigroup $e^{-sA}$ acts separately on each
configuration component, with the Mehler kernel
\[
 m_{s,\omega}(q,r)
 =\left(\frac{e^{s\omega}}{2\pi\sinh(s\omega)}\right)^{N/2}
 \exp\left\{-\frac{\cosh(s\omega)(|q|^2+|r|^2)-2q\cdot r}
 {2\sinh(s\omega)}\right\}>0.
\]
In particular, for each hole position $h\in\Lambda$,
$e^{-sA}$ is positivity improving on the corresponding
phonon space $L^2(\mathbb R^N,dq)$ for every $s>0$.

Since $J_M$ is bounded, the Dyson expansion is valid in
operator norm:
\[
\begin{aligned}
 e^{-\beta\mathcal L_{\omega,M}^{\rm nd}}
 ={}&e^{-\beta Z|t|}\sum_{m=0}^{\infty}
 \int_{0<s_1<\cdots<s_m<\beta}
 e^{-(\beta-s_m)A}J_M
 e^{-(s_m-s_{m-1})A}\\
 &\hspace{30mm}\cdots J_M e^{-s_1A}\,
 ds_1\cdots ds_m.
\end{aligned}
\]
For $m=0$, the integral denotes $e^{-\beta A}$; the norm of
the $m$-th term is at most $\beta^m\|J_M\|^m/m!$.
Every term is positivity preserving.

Let $0\ne\psi,\varphi\in\mathscr P_M$. Choose configurations
$c_{\rm i},c_{\rm f}$ with $\psi_{c_{\rm i}}\ne0$ and
$\varphi_{c_{\rm f}}\ne0$. If they coincide, the $m=0$ term
has strictly positive matrix element. Otherwise, the Nagaoka
connectivity condition provides a path
\[
 c_{\rm i}=c_0\to c_1\to\cdots\to c_m=c_{\rm f}.
\]
In the $m$-th Dyson term, select the product of hopping operators
corresponding to this path. For all times in the interior of the
integration simplex, each intervening Mehler semigroup has a
strictly positive kernel, and each Weyl translation preserves
strict positivity. Since $|t|>0$, this path gives a strictly
positive contribution to the matrix element with $\varphi$.
All other contributions are nonnegative. Therefore,
\[
 \langle\varphi,e^{-\beta\mathcal L_{\omega,M}^{\rm nd}}
 \psi\rangle>0
 \qquad(\beta>0,\quad 0\ne\varphi,\psi\in\mathscr P_M).
\]
The self-dual cone, positivity improvement, self-adjointness,
and existence of a lowest eigenvalue allow the
Perron--Frobenius--Faris simplicity criterion to be applied.
The lowest eigenvalue in each fixed-$M$ sector is thus simple;
 see \cite{Faris1972} and the related
semigroup method in \cite{Miyao2017NagaokaHH}.
\end{proof}

\begin{prop}[Finite-frequency stability in Case 2]
\label{prop:other-KSV-case2}
Suppose $F=0$, $X=t\ne0$, and
\begin{equation}\label{eq:other-case2-condition}
 \widetilde U^{(2)}>\lambda_{\Lambda,\omega}.
\end{equation}
Assume in addition the Nagaoka connectivity condition just stated.
Then
\[
 \min\sigma(H_{\rm KSVH}(\omega))
 =E_{\rm ref}+\lambda_{\Lambda,\omega}-E_{\rm p}(|\Lambda|-1),
\]
and its ground-state space is a single irreducible
spin-$S_{\max}$ multiplet of dimension $|\Lambda|$.
\end{prop}
\begin{proof}
For $F=0$, \eqref{eq:other-global-comparison} and
\eqref{eq:other-case2-condition} exclude doublons from every ground
state. Since a fully polarized scalar-polaron ground state attains
the lower bound, the claimed ground energy is exact.
Moreover, by \eqref{eq:other-KSV-remainder-kernel}, the no-doublon
space $\mathrm{ran} \Pi_0 $ reduces $K_\omega$, and
\[
 K_\omega\upharpoonright_{\operatorname{ran}\Pi_0}
 =E_{\rm ref}\mathbbm1+\mathcal L_\omega^{\rm nd}.
\]

By Lemma~\ref{lem:other-KSV-fixed-M-PF}, the lowest
eigenvalue of $\mathcal L_\omega^{\rm nd}$ is simple in every
fixed-$M$ sector. It remains to identify its value; simplicity
alone does not imply that these energies agree for different $M$.

Let $\Phi_+$ be the fully polarized scalar-polaron ground state.
It satisfies $K_\omega\Phi_+=(E_{\rm ref}+\lambda_{\Lambda,\omega})
\Phi_+$ and has spin $S_{\max}$. Since $K_\omega$ commutes with
$S_{\rm tot}^-$, the vectors
\[
 \Phi_k=(S_{\rm tot}^-)^k\Phi_+,
 \qquad k=0,\ldots,2S_{\max},
\]
are eigenvectors with the same energy. They are nonzero by the
spin-$S_{\max}$ ladder relations and belong to the successive
fixed-$M$ sectors. The global bound
\eqref{eq:other-global-comparison}, together with
$\widetilde U^{(2)}>\lambda_{\Lambda,\omega}$, shows that
$E_{\rm ref}+\lambda_{\Lambda,\omega}$ is the lowest energy in
every such sector. Since all ground states are doublon-free,
and the no-doublon restriction is exactly
$E_{\rm ref}+\mathcal L_\omega^{\rm nd}$, the simplicity from
Lemma~\ref{lem:other-KSV-fixed-M-PF} implies that each $\Phi_k$
spans its fixed-$M$ ground-state space. Thus no ground-state
multiplets of lower spin occur, and the entire ground-state
space is the single irreducible spin-$S_{\max}$ multiplet
generated by $\Phi_+$.
Finally, undoing the Lang--Firsov transformation preserves total
spin and restores the scalar shift $-E_{\rm p}(|\Lambda|-1)$.
\end{proof}

\begin{cor}[Volume-uniform stability in the $X=t$ branches]
\label{cor:other-KSV-large-frequency}
Assume $F\ge0$, $X=t$, and
\begin{equation}\label{eq:other-KSV-positive-condition}
 \Delta_{\rm KSV,\infty}^{(2)}(E_{\rm p})
 :=U-2E_{\rm p}
 -Z\left(2|t|+\left|V-\frac F2-2|t|\right|+|F'|\right)>0.
\end{equation}
If $F>0$, all ground states have maximum spin, with the
multiplicities in Proposition~\ref{prop:other-KSV-case1b}.
If $F=0$, assume $t\ne0$ and the Nagaoka connectivity condition;
then the ground space is the single multiplet of
Proposition~\ref{prop:other-KSV-case2}.
For $|t|E_{\rm p}>0$, the conclusions hold for every
\[
 \omega\ge\omega_0^{(2)}
 :=\frac{2Z|t|E_{\rm p}}{\Delta_{\rm KSV,\infty}^{(2)}(E_{\rm p})}.
\]
If $|t|E_{\rm p}=0$, they hold for every $\omega>0$.
The threshold is independent of $|\Lambda|$ for fixed $Z$ and
fixed model parameters.
\end{cor}
\begin{proof}
Here $\widetilde U^{(2)}=\Delta_{\rm KSV,\infty}^{(2)}(E_{\rm p})$.
By Lemma~\ref{lem:lambda},
$\lambda_{\Lambda,\omega}\le Z|t|E_{\rm p}/\omega$.
If $|t|E_{\rm p}>0$ and $\omega\ge\omega_0^{(2)}$, then
\[
 \lambda_{\Lambda,\omega}
 \le\frac{\Delta_{\rm KSV,\infty}^{(2)}(E_{\rm p})}{2}
 <\widetilde U^{(2)}.
\]
If $|t|E_{\rm p}=0$, then
$\lambda_{\Lambda,\omega}=0<\widetilde U^{(2)}$
for every $\omega>0$.
The conclusions now follow from
Proposition~\ref{prop:other-KSV-case1b} for $F>0$
and Proposition~\ref{prop:other-KSV-case2} for $F=0$.
\end{proof}

\begin{rem}
The strictly positive condition
\eqref{eq:other-KSV-positive-condition} is sufficient for the
Holstein stability assertions above.  No assertion is made at the
boundary $\widetilde U^{(2)}=0$, even though the purely electronic
KSV Case~1(b) admits a boundary formulation.
\end{rem}

%============================================================
\section{Proof of stability under a quantized radiation field}
\label{app:radiation}
%============================================================
Throughout this appendix, we work in the gauge $t\le0$, as in
the hopping-sign convention in Subsection~\ref{subsec:background-KSV-one-hole-ferromagnetism}, and use the one-hole
sector.  We first establish Theorem~\ref{thm:KSV-radiation} and
Corollary~\ref{cor:KSV-radiation-other}, then explain the
simultaneous coupling in
Remark~\ref{rem:simultaneous-boson-couplings}.

\subsection{Photon Fock space factorization and ground-state existence}

The transverse vector potential $A_L(r)$ and the
bond fields $\mathcal A_{xy}$ are defined in
Subsection~\ref{subsec:main-result-radiation};
see \eqref{eq:radiation-field}.
We first establish the ground-state existence and
the decomposition into coupled and free photon modes.

For fixed $L$, $H_{\rm f}$ has compact resolvent.
Indeed, $\nu_L(k)\ge\nu_{\min}(L)
=\min\{m_0,2\pi/L\}>0$.
For any $E>0$, the total photon number is bounded by
$E/\nu_{\min}(L)$ on the spectral subspace
$\mathbbm{1}_{[0,E]}(H_{\rm f})\mathcal R_L$.
Moreover, only finitely many modes satisfy $\nu_L(k)\le E$.
Hence this spectral subspace is finite-dimensional,
which implies that $H_{\rm f}$ has compact resolvent.
All remaining interaction
terms in \eqref{eq:photon-Hamiltonian} are bounded for fixed
$\Lambda$, so $H_{\rm KSV,\gamma}$ is self-adjoint on
$\mathrm{dom}(H_{\rm f})$, bounded below, and has compact
resolvent. The same holds for $L_{\Lambda,\gamma}$ and its
fixed-magnetization analogues.

Because of the ultraviolet cutoff and the convention
$\varepsilon(0,\ell)=0$, only finitely many photon modes couple
to the electrons at fixed $L$. Set
\[
 \mathcal K_{L,\kappa}:=\{k\in(2\pi/L)\mathbb Z^3:
 0<|k|\le\kappa\},
\]
and define
\[
 \begin{aligned}
 \mathfrak h_L^{\rm c}
 :=\ell^2(\mathcal K_{L,\kappa}\times\{1,2\}),\qquad
 \mathfrak h_L^{\rm free}
 :=\ell^2\bigl(((2\pi/L)\mathbb Z^3
 \setminus\mathcal K_{L,\kappa})\times\{1,2\}\bigr).
 \end{aligned}
\]
Thus $\mathfrak h_L=\mathfrak h_L^{\rm c}\oplus
\mathfrak h_L^{\rm free}$, where the second subspace
includes the zero-momentum modes. Both subspaces reduce
$\nu_L$.

Let $\Gamma_{\rm s}(\mathfrak h)$ denote the bosonic Fock
space over a Hilbert space $\mathfrak h$.
The standard Fock space factorization gives
\begin{equation}\label{eq:photon-Fock-factorization}
 \mathcal R_L\simeq\mathcal R_L^{\rm c}\otimes
 \mathcal R_L^{\rm free},\qquad
 \mathcal R_L^{\rm c}:=\Gamma_{\rm s}(\mathfrak h_L^{\rm c}),\quad
 \mathcal R_L^{\rm free}:=\Gamma_{\rm s}(\mathfrak h_L^{\rm free}).
\end{equation}
Correspondingly, the free radiation Hamiltonian decomposes as
\[
 H_{\rm f}=H_{\rm f}^{\rm c}\otimes\mathbbm1
 +\mathbbm1\otimes H_{\rm f}^{\rm free},
\]
where
\[
 \begin{aligned}
 H_{\rm f}^{\rm c}
 &=\sum_{k\in\mathcal K_{L,\kappa}}\sum_{\ell=1}^2
 \nu_L(k)a^*(k,\ell)a(k,\ell),\\
 H_{\rm f}^{\rm free}
 &=\sum_{k\notin\mathcal K_{L,\kappa}}\sum_{\ell=1}^2
 \nu_L(k)a^*(k,\ell)a(k,\ell).
 \end{aligned}
\]
Thus $H_{\rm f}^{\rm c}$ describes the  energy of the
photon modes that may couple to the electrons, while
$H_{\rm f}^{\rm free}$ describes the remaining uncoupled modes,
including the zero-momentum modes.
All $\mathcal A_{xy}$ act trivially on
$\mathcal R_L^{\rm free}$. Thus, by
\eqref{eq:photon-Fock-factorization},
\begin{equation}\label{eq:photon-factorization}
 \begin{aligned}
 H_{\rm KSV,\gamma}
 &=H_{\rm KSV,\gamma}^{\rm c}\otimes\mathbbm1
   +\mathbbm1\otimes H_{\rm f}^{\rm free},\\
 L_{\Lambda,\gamma}
 &=L_{\Lambda,\gamma}^{\rm c}\otimes\mathbbm1
   +\mathbbm1\otimes H_{\rm f}^{\rm free}.
 \end{aligned}
\end{equation}
Here $H_{\rm KSV,\gamma}^{\rm c}$ and
$L_{\Lambda,\gamma}^{\rm c}$ act on
$\mathcal H_\Lambda^{\rm 1h}\otimes\mathcal R_L^{\rm c}$
and $\ell^2(\Lambda)\otimes\mathcal R_L^{\rm c}$,
respectively. They contain all terms of the corresponding
full Hamiltonians except $H_{\rm f}^{\rm free}$,
with $H_{\rm f}^{\rm c}$ as the free-field Hamiltonian
on $\mathcal R_L^{\rm c}$.
Since every free photon mode has strictly positive
frequency, $H_{\rm f}^{\rm free}\ge0$ has the unique ground
state $\Omega_{\rm free}$. Consequently,
$\lambda_{\Lambda,\gamma}=\min\sigma(L_{\Lambda,\gamma}^{\rm c})$,
and the ground-state spaces of $H_{\rm KSV,\gamma}$ and
$L_{\Lambda,\gamma}$ are obtained by tensoring those of
their coupled-mode counterparts with $\Omega_{\rm free}$.
For positivity arguments, it therefore suffices to work
on the finite-mode space $\mathcal R_L^{\rm c}$.
The Schr\"odinger representation and its positivity
properties are established in
Lemma~\ref{lem:photon-Gaussian} at the end of this appendix,
using the same rotation idea as in
\cite[Section~6]{Miyao2017NagaokaHH}.

\begin{rem}
The compact-support assumption on the ultraviolet cutoff
$\chi_\kappa$ is used to reduce the photon field to finitely
many modes coupled to the electrons, allowing us to work
in the usual finite-dimensional Schr\"odinger representation.
This assumption is not essential to the comparison method.
For cutoffs with noncompact support, such as Gaussian cutoffs,
one can instead use the infinite-dimensional Gaussian
$Q$-representation of the photon Fock space.
The same positivity arguments apply, provided the corresponding
photon form factors are square integrable and their vacuum
variances admit a volume-uniform bound.
\end{rem}

\subsection{KSV positive decomposition with Peierls coupling}
Let $\Phi=\Phi_{xy}^\gamma$ for a fixed oriented bond and use
$\mathfrak d=|X-t|$, $s=\operatorname{sgn}(X-t)$, and
$r=\operatorname{sgn}(\widetilde F'(a))$.  Conjugating the
local factors \eqref{eq:PQ} and \eqref{eq:AB} by
$W_{xy}^\gamma$ gives
\begin{equation}\label{eq:photon-covariant-factors}
\begin{aligned}
 P_{xy\sigma}^\gamma&=(1-n_{x,-\sigma})
 (c_{x\sigma}-\Phi c_{y\sigma})(1-n_{y,-\sigma}),\\
 Q_{xy\sigma}^\gamma&=n_{x,-\sigma}
 (c_{x\sigma}-\Phi c_{y\sigma})n_{y,-\sigma},\\
 A_{xy}^\gamma&=a^{-1}(c_{x\downarrow}c_{x\uparrow}
 +\Phi^2c_{y\downarrow}c_{y\uparrow})
 +as\Phi(c_{y\downarrow}c_{x\uparrow}
 +c_{x\downarrow}c_{y\uparrow}),\\
 B_{xy}^\gamma&=c_{x\downarrow}c_{x\uparrow}
 +r\Phi^2c_{y\downarrow}c_{y\uparrow}.
\end{aligned}
\end{equation}
The choice of $r$ is immaterial when $\widetilde F'(a)=0$.
Since $W_{xy}^\gamma$ commutes with all electronic number
operators $n_{z\sigma}$ and local electron spin operators
$\bs S_z$, conjugating the CAR identities in
Appendix~\ref{app:CAR} by $W_{xy}^\gamma$ yields
the corresponding identities with Peierls phases.
 For example, with $T_{xy}^\gamma$ and
$C_{xy}^\gamma$ denoting the Peierls-dressed ordinary and
correlated hopping terms, and with $J_{xy}^\gamma$ the
dressed pair hopping term, they read
\begin{equation}\label{eq:photon-local-identities}
\begin{aligned}
 \sum_\sigma\bigl(P_{xy\sigma}^\gamma
 (P_{xy\sigma}^\gamma)^*
 +(Q_{xy\sigma}^\gamma)^*Q_{xy\sigma}^\gamma\bigr)&\\
 &=4\mathbbm1-3(n_x+n_y)+2(d_x+d_y)
 +2n_xn_y+T_{xy}^\gamma-C_{xy}^\gamma,\\
 \mathfrak d(A_{xy}^\gamma)^*A_{xy}^\gamma
 &=\frac{\mathfrak d}{a^2}(d_x+d_y+J_{xy}^\gamma)
 +a^2\mathfrak d\left(\frac12n_xn_y
 -2\bs S_x\cdot\bs S_y\right)+(X-t)C_{xy}^\gamma,\\
 |\widetilde F'(a)|(B_{xy}^\gamma)^*B_{xy}^\gamma
 &=|\widetilde F'(a)|(d_x+d_y)
 +\widetilde F'(a)J_{xy}^\gamma.
\end{aligned}
\end{equation}
Ordinary and correlated hopping acquire a single Peierls
factor, whereas pair hopping acquires its square.
Since the Peierls coupling is introduced separately on each
bond, we conjugate each local KSV identity by the corresponding
unitary and then sum over the bonds.
Unitary conjugation preserves the positivity of each local
square term. The diagonal terms remain unchanged, so the
graph inequalities used in \eqref{eq:B} continue to hold.

Put
\[
 \mathcal T_{+,\gamma}:=|t|\sum_{\{x,y\}\in\mathscr B_\Lambda}
 \sum_\sigma\bigl(P_{xy\sigma}^\gamma
 (P_{xy\sigma}^\gamma)^*+
 (Q_{xy\sigma}^\gamma)^*Q_{xy\sigma}^\gamma\bigr).
\]
The remainder $\mathcal B_\gamma$ is obtained from
\eqref{eq:B} by replacing $A_{xy},B_{xy}$ with
$A_{xy}^\gamma,B_{xy}^\gamma$, leaving the diagonal
terms unchanged.
Thus $\mathcal B_\gamma\ge0$, and
\begin{equation}\label{eq:photon-dressed-KSV}
 H_{\rm KSV,\gamma}=E_{\rm ref}\mathbbm1
 +\mathcal L_\gamma+\widetilde U_{\rm rad}(a)\mathscr D
 +\widetilde F(a)\mathscr S+\mathcal B_\gamma,
 \qquad\mathcal L_\gamma:=H_{\rm f}+\mathcal T_{+,\gamma}\ge0.
\end{equation}
The operators $\mathcal B_\gamma$, $\mathscr D$, and
$\mathscr S$ vanish on the entire maximum-spin subspace.
Conjugating the hopping identity
\eqref{eq:local-kinetic-identity} by $W_{xy}^\gamma$
introduces the Peierls phases into the hopping terms,
while leaving the electron number factors unchanged.
Consequently, the retained kinetic operator still
commutes with the doublon-number operator $\mathscr D$.
 As in Lemma~\ref{lem:doublon},
\begin{equation}\label{eq:photon-charge-commutation}
 [\mathcal T_{+,\gamma},\mathscr D]=0,
 \qquad[\mathcal L_\gamma,\Pi_0]=0.
\end{equation}

\subsection{Energy comparison in the no-doublon sector}
By the CAR calculation in Lemma~\ref{lem:no-doublon-polaron},
an electron hop from $y$ into a hole at $x$ carries
the fermionic sign $-1$ and transfers the spin at $y$
to $x$ through the unitary $U_{yx}:\mathcal S_x\to\mathcal S_y$.
The Peierls coupling introduces the additional factor
$\Phi_{xy}^\gamma$, acting only on the photon Fock space.
Consequently,
\begin{equation}\label{eq:photon-spin-fiber-operator}
 \mathcal L_\gamma^{\rm nd}=H_{\rm f}+Z|t|\mathbbm1
 -|t|\sum_{(x,y)\in\overrightarrow{\mathscr B}_\Lambda}
 |y\rangle\langle x|\otimes U_{yx}
 \otimes\Phi_{xy}^\gamma.
\end{equation}
Moreover, the scalar Hamiltonian \eqref{eq:photon-scalar}
is nonnegative, since its form is
\begin{equation}\label{eq:photon-squares}
 \mathfrak q_{L_{\Lambda,\gamma}}[\varphi]
 =\mathfrak q_{H_{\rm f}}[\varphi]
 +|t|\sum_{\{x,y\}\in\mathscr B_\Lambda}
 \|\varphi_y-\Phi_{xy}^\gamma\varphi_x\|^2.
\end{equation}
Each unoriented bond is counted exactly once.

\begin{lem}[Lower bound by the scalar one-hole ground-state energy]
\label{lem:photon-modulus}
For the standard radiation field \eqref{eq:radiation-field},
\begin{equation}\label{eq:photon-modulus}
 \mathcal L_\gamma^{\rm nd}
 \ge\lambda_{\Lambda,\gamma}\mathbbm1
\end{equation}
in the sense of closed quadratic forms.
\end{lem}

\begin{proof}
By \eqref{eq:photon-Fock-factorization} and
\eqref{eq:photon-spin-fiber-operator},
$\mathcal L_\gamma^{\rm nd}$ decomposes into the
coupled-mode operator $\mathcal L_{\gamma,{\rm c}}^{\rm nd}$
and the nonnegative free-photon Hamiltonian
$H_{\rm f}^{\rm free}$. Thus, to prove
\eqref{eq:photon-modulus}, it suffices to establish
\[
 \mathcal L_{\gamma,{\rm c}}^{\rm nd}
 \ge\lambda_{\Lambda,\gamma}\mathbbm1
\]
as quadratic forms on
$\bigoplus_{x\in\Lambda}
(\mathcal S_x\otimes\mathcal R_L^{\rm c})$.
By \eqref{eq:photon-factorization} and the uniqueness of
the ground state of $H_{\rm f}^{\rm free}$,
\[
 \lambda_{\Lambda,\gamma}
 =\min\sigma(L_{\Lambda,\gamma}^{\rm c}).
\]

Let $m=\dim\mathfrak h_L^{\rm c}$.
By Lemma~\ref{lem:photon-Gaussian}, we identify
$\mathcal R_L^{\rm c}$ unitarily with
$L^2(\mathbb R^m,dq)$.
For $\Psi\in\bigoplus_{x\in\Lambda}
(\mathcal S_x\otimes\mathcal R_L^{\rm c})$,
let $\widehat\Psi$ be the image of $\Psi$ under
the field rotation $\xi$ followed by the unitary
identification $\mathcal R_L^{\rm c}\simeq
L^2(\mathbb R^m,dq)$ of Lemma~\ref{lem:photon-Gaussian}.
 Writing
$\widehat\Psi_x(q)\in\mathcal S_x$ for its component
at hole position $x$, define
\[
 \rho_x(q):=\|\widehat\Psi_x(q)\|_{\mathcal S_x}.
\]
Then $\rho\in\ell^2(\Lambda)\otimes L^2(\mathbb R^m,dq)$
and $\|\rho\|=\|\widehat\Psi\|=\|\Psi\|$.

By \eqref{eq:photon-positive-shift}, the rotated Peierls
operators act as real translations. Together with the
unitarity of the spin transport $U_{yx}$, pointwise
Cauchy--Schwarz gives
\begin{equation}\label{eq:photon-hop-modulus}
 \operatorname{Re}\langle\widehat\Psi_y,
 (U_{yx}\otimes\widehat\Phi_{xy}^\gamma)
 \widehat\Psi_x\rangle
 \le\langle\rho_y,\widehat\Phi_{xy}^\gamma\rho_x\rangle.
\end{equation}
The hopping operators are bounded, so this step requires
no additional domain assumption.

For the free-photon part, put
$S_s=e^{-sH_{\rm f}^{\rm c}}$ in the same
Schr\"odinger representation.
For every $s>0$, the Mehler kernel of $S_s$ is
strictly positive, so $S_s$ is positivity improving
on $L^2(\mathbb R^m,dq)$.
On each spin fiber, $S_s$ acts as the identity
on the spin variables. Hence
\[
 \|(S_s\widehat\Psi)_x(q)\|_{\mathcal S_x}
 \le(S_s\rho_x)(q)
\]
for almost every $q\in\mathbb R^m$, which implies
\[
 \operatorname{Re}\langle\widehat\Psi,S_s\widehat\Psi\rangle
 \le\langle\rho,S_s\rho\rangle.
\]
Since $\|\rho\|=\|\widehat\Psi\|$, for $s>0$,
\[
 \frac1s\langle\rho,(\mathbbm1-S_s)\rho\rangle
 \le\frac1s\langle\widehat\Psi,(\mathbbm1-S_s)
 \widehat\Psi\rangle.
\]
The spectral theorem, using
$(1-e^{-sE})/s\uparrow E$ as $s\downarrow0$, shows
without assuming it in advance that
$\rho\in\mathcal Q(H_{\rm f}^{\rm c})$ and
\[
 \mathfrak q_{H_{\rm f}^{\rm c}}[\rho]
 \le\mathfrak q_{H_{\rm f}^{\rm c}}[\widehat\Psi]
 =\mathfrak q_{H_{\rm f}^{\rm c}}[\Psi].
\]
Here the last equality follows because $\xi$
commutes with $H_{\rm f}^{\rm c}$.

Combining this with \eqref{eq:photon-hop-modulus} gives
\[
 \mathfrak q_{\mathcal L_{\gamma,{\rm c}}^{\rm nd}}[\Psi]
 \ge\mathfrak q_{\xi L_{\Lambda,\gamma}^{\rm c}\xi^*}[\rho]
 \ge\lambda_{\Lambda,\gamma}\|\Psi\|^2.
\]
Finally, by the Fock space decomposition
\eqref{eq:photon-Fock-factorization},
\[
 \mathcal L_\gamma^{\rm nd}
 =\mathcal L_{\gamma,{\rm c}}^{\rm nd}\otimes\mathbbm1
 +\mathbbm1\otimes H_{\rm f}^{\rm free}
 \ge\lambda_{\Lambda,\gamma}\mathbbm1.
\]
Thus \eqref{eq:photon-modulus} holds on the full
form domain.
\end{proof}

\subsection{Energy comparison and all three KSV branches}
\begin{proof}[Proof of Theorem~\ref{thm:KSV-radiation}]
By \eqref{eq:photon-charge-commutation}, $\Pi_0$ reduces
$\mathcal L_\gamma$. Lemma~\ref{lem:photon-modulus}
gives the lower bound $\lambda_{\Lambda,\gamma}$ on
$\operatorname{ran}\Pi_0$, while $\mathcal L_\gamma\ge0$
on $\operatorname{ran}(1-\Pi_0)$. Consequently,
\[
 \mathcal L_\gamma-\lambda_{\Lambda,\gamma}\Pi_0\ge0,
\]
as in \eqref{eq:L-lower} for the Holstein case.

Since $\mathscr D\ge\Pi_+$ and $\lambda_{\Lambda,\gamma}\ge0$,
\eqref{eq:photon-dressed-KSV} gives
\begin{equation}\label{eq:photon-global-identity}
\begin{aligned}
 &H_{\rm KSV,\gamma}-(E_{\rm ref}+\lambda_{\Lambda,\gamma})
 \mathbbm1
 -(\widetilde U_{\rm rad}(a)-\lambda_{\Lambda,\gamma})
 \mathscr D-\widetilde F(a)\mathscr S\\
 &\quad=(\mathcal L_\gamma-\lambda_{\Lambda,\gamma}\Pi_0)
 +\mathcal B_\gamma
 +\lambda_{\Lambda,\gamma}(\mathscr D-\Pi_+)\ge0.
\end{aligned}
\end{equation}
This is the radiation-field counterpart of
\eqref{eq:exact-positive-difference}.

As in the proof of Proposition~\ref{prop:energy} for the
Holstein case, we establish that the lower bound is attained
by embedding a scalar ground-state eigenvector into the
fully spin-up one-hole sector.
Let $\varphi$ be a normalized ground-state eigenvector of
$L_{\Lambda,\gamma}$, so that
\[
 L_{\Lambda,\gamma}\varphi
 =\lambda_{\Lambda,\gamma}\varphi.
\]
Embed $\varphi$ into the fully spin-up one-hole sector.
On this sector, $\mathcal L_\gamma$ agrees with
$L_{\Lambda,\gamma}$, while $\mathcal B_\gamma$,
$\mathscr D$, and $\mathscr S$ vanish. Hence the embedded
state attains the lower bound in
\eqref{eq:photon-global-identity}.

Under the strict coefficient conditions, every ground
state $\Psi$ therefore satisfies
$\mathscr D\Psi=\mathscr S\Psi=0$.
Repeating the spin-alignment argument in the proof of
Proposition~\ref{prop:spin}, we conclude that
$\Psi$ has total spin $S_{\max}$.

For $t\ne0$, we prove that the lowest eigenvalue of
$L_{\Lambda,\gamma}$ is simple, following the
positivity-improving argument for the Holstein polaron in
Proposition~\ref{prop:PF}.
By \eqref{eq:photon-Fock-factorization} and
\eqref{eq:photon-factorization},
\[
 L_{\Lambda,\gamma}
 =L_{\Lambda,\gamma}^{\rm c}\otimes\mathbbm1
 +\mathbbm1\otimes H_{\rm f}^{\rm free}.
\]
Since $H_{\rm f}^{\rm free}$ has the unique ground state
$\Omega_{\rm free}$, it suffices to prove simplicity
of the lowest eigenvalue of $L_{\Lambda,\gamma}^{\rm c}$
on $\ell^2(\Lambda)\otimes\mathcal R_L^{\rm c}$.

Let $m=\dim\mathfrak h_L^{\rm c}$.
By Lemma~\ref{lem:photon-Gaussian}, we identify
$\mathcal R_L^{\rm c}$ unitarily with
$L^2(\mathbb R^m,dq)$.
Under this identification, the field-rotated operator
$\xi L_{\Lambda,\gamma}^{\rm c}\xi^*$ acts on
\[
 \ell^2(\Lambda)\otimes L^2(\mathbb R^m,dq)
 \simeq\bigoplus_{x\in\Lambda}L^2(\mathbb R^m,dq),
\]
where $\xi$ acts trivially on the hole factor.
We equip this space with the cone
\[
 \mathscr P_{\rm rad}:=
 \left\{f=(f_x)_{x\in\Lambda}:
 f_x(q)\ge0\ \text{for a.e. }q\in\mathbb R^m
 \text{ and every }x\in\Lambda\right\}.
\]
Since $\xi$ commutes with $H_{\rm f}^{\rm c}$,
the rotated operator is
\[
 \xi L_{\Lambda,\gamma}^{\rm c}\xi^*
 =H_{\rm f}^{\rm c}+Z|t|\mathbbm1
 -|t|\sum_{(x,y)\in\overrightarrow{\mathscr B}_\Lambda}
 |y\rangle\langle x|\otimes\widehat\Phi_{xy}^\gamma.
\]
By Lemma~\ref{lem:photon-Gaussian}, each
$\widehat\Phi_{xy}^\gamma$ is a positivity-preserving
translation, and $e^{-sH_{\rm f}^{\rm c}}$ is positivity
improving for every $s>0$.

The hopping operator is bounded and positivity preserving
with respect to $\mathscr P_{\rm rad}$.
Consequently, the Dyson expansion of
$e^{-\beta\xi L_{\Lambda,\gamma}^{\rm c}\xi^*}$
converges in operator norm, and every term is
positivity preserving.
To prove positivity improvement, take nonzero
$f,g\in\mathscr P_{\rm rad}$ and choose hole positions
$x,y$ such that $f_x\ne0$ and $g_y\ne0$.
Since $\Lambda$ is connected, there is a path of
allowed hole moves from $x$ to $y$.
The Dyson term corresponding to this path has a
strictly positive matrix element between $f$ and $g$:
the Peierls translations preserve positivity, while
the intervening free photon semigroups improve it.
All other Dyson terms are nonnegative.
Thus
\[
 \langle g,e^{-\beta\xi L_{\Lambda,\gamma}^{\rm c}\xi^*}
 f\rangle>0
\]
for every $\beta>0$.
Since $L_{\Lambda,\gamma}^{\rm c}$ has compact resolvent,
the Perron--Frobenius--Faris theorem implies that its
lowest eigenvalue is simple.

By \eqref{eq:photon-factorization}, the uniqueness of
$\Omega_{\rm free}$ then implies that the lowest
eigenvalue of $L_{\Lambda,\gamma}$ is also simple.
Finally, the construction of the maximum-spin intertwiner
in Proposition~\ref{prop:intertwiner} applies to
$H_{\rm KSV,\gamma}$, since the Peierls coupling preserves
spin rotation invariance and acts identically on every
spin weight. Replacing $L_{{\rm pol},\omega}$ by
$L_{\Lambda,\gamma}$ in the corresponding unitary
equivalence shows that the ground-state space is a
single spin-$S_{\max}$ multiplet.

If $t=0$, the hole cannot move, and each of the $|\Lambda|$
hole positions supports an independent photon vacuum and
a spin-$S_{\max}$ multiplet of dimension $|\Lambda|$.
The total ground-state degeneracy is therefore $|\Lambda|^2$.

For the uniform statement, let
$c=\mathfrak d^2/F$ and choose
$a_\theta^2=(1-\theta)F/\mathfrak d$.
The coefficient estimate in Lemma~\ref{lem:a-choice}, now with
$E_{\rm p}=0$, gives, for $0<\theta\le1/2$,
\[
 \widetilde F(a_\theta)=\theta F>0,\qquad
 |\widetilde U_{\rm rad}(a_\theta)-\Delta_{\rm rad}^{(1)}|
 \le C\theta,\quad C=Z(F/2+4c).
\]
The variational bound \eqref{eq:photon-lambda-bounds},
established in Subsection~\ref{subsec:photon-uniform-bound},
gives $\lambda_{\Lambda,\gamma}\le\eta_\gamma$.  If
$\delta=\Delta_{\rm rad}^{(1)}-\eta_\gamma>0$, choose
$0<\theta\le\min\{1/4,\delta/(4C)\}$. Then
\[
 \widetilde U_{\rm rad}(a_\theta)
 \ge\eta_\gamma+3\delta/4>\lambda_{\Lambda,\gamma}.
\]
This choice is independent of both volume parameters.
If $\Delta_{\rm rad}^{(1)}>Z|t|$, the same argument, using
$\lambda_{\Lambda,\gamma}\le Z|t|$, gives an $a$ valid
for every $\gamma\in\mathbb R$.
\end{proof}

\begin{proof}[Proof of Corollary~\ref{cor:KSV-radiation-other}]
For $X=t$, we adapt the positive-decomposition argument
for the Holstein branches in
Appendix~\ref{app:other-KSV-branches}.
Starting from \eqref{eq:other-KSV-identity} in
Lemma~\ref{lem:other-KSV-decomposition} with $E_{\rm p}=0$,
we apply the bondwise Peierls conjugations described in
\eqref{eq:photon-local-identities}.
The resulting coefficient of $\mathscr D$ is
$\Delta_{\rm rad}^{(2)}$.
The corresponding remainder $\mathcal B_\gamma^{(2)}$
is obtained from $\mathcal B^{(2)}$ in
\eqref{eq:other-KSV-remainder} by conjugating the local
square terms bondwise and leaving the diagonal terms
unchanged.
The square terms remain positive and annihilate the
no-doublon subspace, while the diagonal terms retain
their original properties. Consequently,
$\mathcal B_\gamma^{(2)}\ge0$ and
\[
 \mathcal B_\gamma^{(2)}\Pi_0
 =\Pi_0\mathcal B_\gamma^{(2)}=0.
\]
Since the retained operator $\mathcal L_\gamma$ is unchanged,
Lemma~\ref{lem:photon-modulus} gives
\[
 \mathcal L_\gamma^{\rm nd}
 \ge\lambda_{\Lambda,\gamma}\mathbbm1.
\]
Thus, exactly as in \eqref{eq:other-global-comparison}
for the Holstein case, we obtain
\[
 H_{\rm KSV,\gamma}-(E_{\rm ref}+\lambda_{\Lambda,\gamma})
 \mathbbm1
 \ge(\Delta_{\rm rad}^{(2)}-\lambda_{\Lambda,\gamma})
 \mathscr D+F\mathscr S.
\]
For $F>0$, the proof of
Proposition~\ref{prop:other-KSV-case1b} applies with
$L_{{\rm pol},\omega}$ replaced by $L_{\Lambda,\gamma}$.
In particular, the positive spin-defect term $F\mathscr S$
forces every ground state to have maximum spin.

Now suppose $F=0$, $X=t\ne0$, and the Nagaoka connectivity
condition holds. We adapt the fixed-magnetization argument of
Proposition~\ref{prop:other-KSV-case2} to the radiation-field model.
As established above, the Peierls-dressed remainder satisfies
$\mathcal B_\gamma^{(2)}\Pi_0
=\Pi_0\mathcal B_\gamma^{(2)}=0$.
Moreover, $[\mathcal L_\gamma,\Pi_0]=0$ and
$\mathscr D\Pi_0=0$.
Since $F=0$, these identities show that $\Pi_0$
reduces $H_{\rm KSV,\gamma}$ and
\[
 H_{\rm KSV,\gamma}\upharpoonright_{\operatorname{ran}\Pi_0}
 =E_{\rm ref}\mathbbm1+\mathcal L_\gamma^{\rm nd},
\]
where $\mathcal L_\gamma^{\rm nd}$ is given by
\eqref{eq:photon-spin-fiber-operator}.

Fix an eigenvalue $M$ of $S_{\rm tot}^{(3)}$, and let
$\mathscr C_M$ be the set of hole--spin configurations
introduced in Appendix~\ref{app:other-KSV-branches}.
Two configurations are adjacent if one is obtained from
the other by moving the hole across a bond.
By the basis \eqref{eq:hole-basis}, the electronic
no-doublon subspace with magnetization $M$ is identified
with $\ell^2(\mathscr C_M)$.

By \eqref{eq:photon-Fock-factorization}, the corresponding
electron--photon Hilbert space is
$
 \ell^2(\mathscr C_M)\otimes\mathcal R_L^{\rm c}
 \otimes\mathcal R_L^{\rm free}.
$
The decomposition \eqref{eq:photon-factorization}
also holds for the restriction to this subspace.
Since $H_{\rm f}^{\rm free}$ has the unique ground state
$\Omega_{\rm free}$, it suffices to prove simplicity
of the lowest eigenvalue of the coupled-mode operator
on $\ell^2(\mathscr C_M)\otimes\mathcal R_L^{\rm c}$.

Let $m=\dim\mathfrak h_L^{\rm c}$.
By Lemma~\ref{lem:photon-Gaussian}, we identify
$\mathcal R_L^{\rm c}$ unitarily with
$L^2(\mathbb R^m,dq)$.
After conjugation by $\mathbbm1\otimes\xi$,
the coupled-mode operator acts on
\[
 \mathcal H_M:=
 \ell^2(\mathscr C_M)\otimes L^2(\mathbb R^m,dq)
 \simeq\bigoplus_{c\in\mathscr C_M}L^2(\mathbb R^m,dq).
\]
We equip this space with the cone
\[
 \mathscr P_M:=
 \left\{\psi=(\psi_c)_{c\in\mathscr C_M}:
 \psi_c(q)\ge0\ \text{for a.e. }q\in\mathbb R^m
 \text{ and every }c\in\mathscr C_M\right\}.
\]

The restriction of $\mathcal L_{\gamma,{\rm c}}^{\rm nd}$
to the fixed-$M$ subspace, after conjugation by
$\mathbbm1\otimes\xi$, is
\[
 L_{\gamma,M}^{\rm nd}
 =H_{\rm f}^{\rm c}+Z|t|\mathbbm1-J_M,
 \qquad
 J_M=|t|\sum_{c\to c'}|c'\rangle\langle c|
 \otimes\widehat\Phi_{h(c),h(c')}^\gamma,
\]
where the sum runs over directed edges of the configuration
graph and $h(c)$ denotes the hole position in $c$.
Here $H_{\rm f}^{\rm c}$ acts on each configuration
component. This is the radiation-field counterpart of
the fixed-magnetization operator in
Lemma~\ref{lem:other-KSV-fixed-M-PF}.

Each configuration has exactly $Z$ allowed hole moves,
and reversing a move replaces the corresponding Peierls
operator by its adjoint. Thus $J_M$ is bounded and
self-adjoint. By Lemma~\ref{lem:photon-Gaussian},
each $\widehat\Phi_{xy}^\gamma$ is a positivity-preserving
unitary on $L^2(\mathbb R^m,dq)$.
Consequently, $J_M$ preserves $\mathscr P_M$.

Put $S_s=e^{-sH_{\rm f}^{\rm c}}$, acting on each configuration
component in the Schr\"odinger representation.
By Lemma~\ref{lem:photon-Gaussian}, $S_s$ has a strictly
positive Mehler kernel for every $s>0$.
Moreover, by \eqref{eq:photon-positive-shift}, each
$\widehat\Phi_{xy}^\gamma$ acts as a real translation
and therefore preserves nonnegative functions.
Thus the free semigroup and the hopping operators
have precisely the positivity properties used in
the proof of Lemma~\ref{lem:other-KSV-fixed-M-PF}.

Since $J_M$ is bounded and preserves $\mathscr P_M$,
the same Dyson expansion and path argument applies.
Indeed, the Nagaoka connectivity condition connects
any two configurations by allowed hole moves, and
the contribution of each such path is strictly positive
because the intervening free semigroups are positivity
improving. Consequently,
\[
 \langle\varphi,e^{-\beta L_{\gamma,M}^{\rm nd}}\psi\rangle>0
 \qquad
 (\beta>0,\quad 0\ne\varphi,\psi\in\mathscr P_M).
\]
Hence $e^{-\beta L_{\gamma,M}^{\rm nd}}$ is positivity
improving with respect to $\mathscr P_M$.

The operator $L_{\gamma,M}^{\rm nd}$ has compact resolvent,
so its lowest spectral value is an eigenvalue.
Perron--Frobenius--Faris therefore implies that it is simple.
The full fixed-$M$ block is obtained by adding the independent
$H_{\rm f}^{\rm free}$, whose vacuum is the unique ground state;
its lowest eigenvalue is therefore simple as well.

Finally, the same $\mathrm{SU}(2)$ argument as in the proof of
Proposition~\ref{prop:other-KSV-case2} applies.
The fully polarized ground state generates a spin-$S_{\max}$
multiplet under the action of $S_{\rm tot}^-$.
By the global energy comparison and the simplicity of the
ground state in each fixed-magnetization sector, this multiplet
exhausts the ground-state space.

Finally, the variational estimate
\eqref{eq:photon-lambda-bounds}, established in
Subsection~\ref{subsec:photon-uniform-bound}, gives
$\lambda_{\Lambda,\gamma}\le\eta_\gamma$.
Thus $\eta_\gamma<\Delta_{\rm rad}^{(2)}$ is sufficient.
If $\Delta_{\rm rad}^{(2)}>Z|t|$, one may instead use
$\lambda_{\Lambda,\gamma}\le Z|t|$, which holds for every
$\gamma\in\mathbb R$.
\end{proof}

\subsection{Uniform photon bound and simultaneous coupling}
\label{subsec:photon-uniform-bound}

\begin{proof}[Finiteness of $v_*$]
For each bond set
\[
 I_{xy}(k,\ell):=\varepsilon(k,\ell)\cdot
 \int_{C_{xy}}e^{\im k\cdot r}\,dr.
\]
Recall that $\varepsilon(0,\ell)=0$, so that
$I_{xy}(0,\ell)=0$. Thus the zero-momentum modes do not
contribute to the bond fields $\mathcal A_{xy}$.
By \eqref{eq:radiation-field} and the oscillator vacuum formula,
\begin{equation}\label{eq:photon-variance-formula}
\begin{aligned}
 v_{xy}&:=\langle\Omega_{\rm f},\mathcal A_{xy}^2
 \Omega_{\rm f}\rangle
 =\frac1{L^3}\sum_{k\ne0,\ell}
 \frac{\chi_\kappa(k)^2|I_{xy}(k,\ell)|^2}
 {2|k|},\\
 \langle\Omega_{\rm f},\Phi_{xy}^\gamma\Omega_{\rm f}\rangle
 &=e^{-\gamma^2v_{xy}/2}.
\end{aligned}
\end{equation}
Since $\operatorname{length}(C_{xy})\le\ell_*$,
we have $|I_{xy}(k,\ell)|\le\ell_*$. Hence
\[
 v_{xy}\le\frac{\|\chi_\kappa\|_\infty^2\ell_*^2}{L^3}
 \sum_{0<|k|\le\kappa}\frac1{|k|}
 \le C\ell_*^2\kappa^2,
\]
where the last estimate uses
$L^{-3}\sum_{0<|k|\le\kappa}|k|^{-1}\le C\kappa^2$
for cubic boxes. This proves the finiteness of $v_*$
claimed in \eqref{eq:photon-variance}.
The uniform hole trial state and the photon vacuum give
\[
 \lambda_{\Lambda,\gamma}
 \le Z|t|-\frac{|t|}{|\Lambda|}
 \sum_{(x,y)\in\overrightarrow{\mathscr B}_\Lambda}
 e^{-\gamma^2v_{xy}/2}\le\eta_\gamma\le Z|t|.
\]
The lower bound follows from \eqref{eq:photon-squares}.
This completes the proof of \eqref{eq:photon-lambda-bounds}.
\end{proof}

\begin{proof}[Proof of Remark~\ref{rem:simultaneous-boson-couplings}]
The Lang--Firsov unitary of \eqref{eq:LF-unitary} commutes
with the photon field. It replaces $U$ with $U-2E_{\rm p}$,
produces the constant $-E_{\rm p}(|\Lambda|-1)$, and dresses
hole hopping with $\mathfrak D_{xy}\otimes\Phi_{xy}^\gamma$.

By \eqref{eq:photon-Fock-factorization}, the photon Fock
space decomposes into coupled and free factors.
For the coupled photon modes, we use the Schr\"odinger
representation of Lemma~\ref{lem:photon-Gaussian}
together with the oscillator coordinates of the Holstein
phonons. After the photon-field rotation, both hopping
factors act as real translations on $L^2(\mathbb R^{N+m})$,
and the free heat semigroup generated by
$\omega N_{\rm ph}+H_{\rm f}^{\rm c}$ has a strictly
positive Mehler kernel.

Thus the pointwise modulus comparison and the semigroup
difference-quotient argument in
Lemma~\ref{lem:photon-modulus} extend to the product space.
The positive-decomposition argument then gives the exact
energy comparison and the maximum-spin conclusion.
The remaining free photon modes contribute only
$H_{\rm f}^{\rm free}\ge0$ and have the unique vacuum
ground state. The ground-state multiplicities follow
from the same positivity-improving arguments as in the
separate-coupling cases.

Finally, using the Holstein and photon vacuum states as
trial states gives
\[
 0\le\lambda_{\Lambda,\omega,\gamma}
 \le Z|t|\left(1-e^{-E_{\rm p}/\omega-\gamma^2v_*/2}\right).
\]
\end{proof}

\subsection{Schr\"odinger representation of the Peierls operators}

For the reader's convenience, we recall the finite-mode
Schr\"odinger representation used in the preceding arguments.
The central point is that the photon-field rotation turns
the Peierls operators into positivity-preserving translations.

\begin{lem}[Schr\"odinger representation and positivity]
\label{lem:photon-Gaussian}
Let $m=\dim\mathfrak h_L^{\rm c}<\infty$.There is a unitary identification
\[
 \mathcal R_L^{\rm c}\simeq L^2(\mathbb R^m,dq),
\]
where $q=(q_1,\ldots,q_m)\in\mathbb R^m$ and $q_j$
denotes the $j$th coordinate.
Under this identification, the photon vacuum is
represented by
\[
 \Omega_{\rm c}(q)=\pi^{-m/4}e^{-|q|^2/2}.
\]
The canonical oscillator operators
\[
 Q_j:=\frac{a_j^*+a_j}{\sqrt2},\qquad
 P_j:=\frac{\im(a_j^*-a_j)}{\sqrt2},
 \quad j=1,\ldots,m,
\]
are essentially self-adjoint on the finite-particle
subspace of $\mathcal R_L^{\rm c}$.
Their self-adjoint closures, transported to
$L^2(\mathbb R^m,dq)$ by the unitary identification,
act on $\mathscr S(\mathbb R^m)$ as
\[
 (Q_jf)(q)=q_jf(q),\qquad
 (P_jf)(q)=-\im\partial_{q_j}f(q).
\]
Every bond field has an expansion
\[
 \mathcal A_{xy}=\sum_{j=1}^m g_{xy,j}Q_j,
 \qquad g_{xy,j}\in\mathbb R,
 \qquad
 \langle\Omega_{\rm f},\mathcal A_{xy}^2\Omega_{\rm f}\rangle
 =\frac12|g_{xy}|^2.
\]
Set $\xi=\exp(\im\pi N_{\rm f}^{\rm c}/2)$,
where $N_{\rm f}^{\rm c}$ is the photon number operator
on $\mathcal R_L^{\rm c}$, and let
$\widehat\Phi_{xy}^\gamma=\xi\Phi_{xy}^\gamma\xi^*$.
Then
\begin{equation}\label{eq:photon-positive-shift}
 (\widehat\Phi_{xy}^\gamma f)(q)=f(q+h_{xy}),
 \qquad h_{xy}=\gamma g_{xy}.
\end{equation}
Thus $\widehat\Phi_{xy}^\gamma$ and its inverse preserve
nonnegative functions. Moreover, $e^{-sH_{\rm f}^{\rm c}}$
is positivity improving on $L^2(\mathbb R^m,dq)$ for every
$s>0$.
\end{lem}

\begin{proof}
Choose a set $K_+$ containing one representative from each
pair $\{k,-k\}$ in $\mathcal K_{L,\kappa}$ and take
$\varepsilon(-k,\ell)=\varepsilon(k,\ell)$.
For $k\in K_+$ and $\ell=1,2$, define
\[
 a_{k\ell,c}
 =\frac{a(k,\ell)+a(-k,\ell)}{\sqrt2},
 \qquad
 a_{k\ell,s}
 =\frac{\im(a(k,\ell)-a(-k,\ell))}{\sqrt2}.
\]
These operators define a unitary change of photon modes
which leaves $H_{\rm f}^{\rm c}$ diagonal.
Label the resulting modes by $j=(k,\ell,\tau)$,
where $\tau=c,s$, and define
\[
 Q_j:=\frac{a_j^*+a_j}{\sqrt2},\qquad
 P_j:=\frac{\im(a_j^*-a_j)}{\sqrt2}.
\]
On the finite-particle subspace of $\mathcal R_L^{\rm c}$,
they satisfy the canonical commutation relations
\[
 [Q_j,P_{j'}]=\im\delta_{jj'}\mathbbm1,\qquad
 [Q_j,Q_{j'}]=[P_j,P_{j'}]=0.
\]
Each $Q_j$ and $P_j$ is essentially self-adjoint on
this subspace, and we use the same notation for their
self-adjoint closures as usual. Their exponentials generate a
regular irreducible representation of the Weyl
commutation relations.
By the Stone--von Neumann uniqueness theorem, there is
a unitary identification
\[
 \mathcal R_L^{\rm c}\simeq L^2(\mathbb R^m,dq),
\]
where $m=\dim\mathfrak h_L^{\rm c}$ and
$q=(q_1,\ldots,q_m)\in\mathbb R^m$.
Under this identification, the self-adjoint closures
of $Q_j$ and $P_j$ act on $\mathscr S(\mathbb R^m)$ as
\[
 (Q_jf)(q)=q_jf(q),\qquad
 (P_jf)(q)=-\im\partial_{q_j}f(q).
\]
The normalized photon vacuum is represented by
$\Omega_{\rm c}(q)=\pi^{-m/4}e^{-|q|^2/2}$.

Every bond field has the expansion
\[
 \mathcal A_{xy}=\sum_jg_{xy,j}Q_j,
\]
where
\[
 g_{xy,k\ell,\tau}
 =\frac{\sqrt2\,\chi_\kappa(k)}{L^{3/2}\sqrt{|k|}}
 \varepsilon(k,\ell)\cdot
 \int_{C_{xy}}f_\tau(k\cdot r)\,dr,
 \qquad f_c=\cos,\quad f_s=\sin.
\]
Consequently,
\[
 \langle\Omega_{\rm f},\mathcal A_{xy}^2\Omega_{\rm f}\rangle
 =\frac12|g_{xy}|^2.
\]

The unitary $\xi$ implements a quarter-turn in the phase
space of each coupled photon oscillator, with
$\xi Q_j\xi^*=P_j$.
Consequently,
\[
 \widehat\Phi_{xy}^\gamma
 =\exp\left(\im\gamma\sum_jg_{xy,j}P_j\right),
\]
which acts as the real translation
\eqref{eq:photon-positive-shift}. The inverse corresponds
to replacing $h_{xy}$ by $-h_{xy}$.
Finally, in the same representation,
\[
 H_{\rm f}^{\rm c}
 =\frac12\sum_{j=1}^m\nu_j
 (-\partial_{q_j}^2+q_j^2-1).
\]
Its free heat semigroup has a strictly positive Mehler kernel
for every $s>0$ and is therefore positivity improving.
The same rotation and positivity mechanisms underlie
\cite[Section~6]{Miyao2017NagaokaHH}.
\end{proof}

\section*{Acknowledgments}
T.M. was supported by JSPS KAKENHI Grant Number JP23H01086.
H.K. was supported by JSPS KAKENHI Grant Number JP23K25790.

\section*{Declaration on the Use of AI Tools}
The problem formulation and the overall direction of the proofs were conceived by the authors. In the subsequent research and preparation of the manuscript, T.M. used ChatGPT Pro (OpenAI) and H.K. used Gemini Pro (Google) to discuss and review mathematical arguments and to improve the exposition of the manuscript. All AI-assisted output was checked by the authors, and the AI tools are not listed as authors. The authors take full responsibility for the mathematical content and conclusions of this paper.

\section*{Conflict of Interest}
The authors declare no conflicts of interest relevant to this article.

\section*{Data Availability}
Data sharing is not applicable to this article, as no datasets
were generated or analyzed during the study.

\bibliographystyle{abbrvurl}
\bibliography{KSVH_Reference_DirectComparison_20261030}
\end{document}